\documentclass[twocolumn]{aastex63}

\usepackage[T1]{fontenc}        
\usepackage{amsmath,mathtools}  
\usepackage{makecell}           
\usepackage{hyperref}           

\hypersetup{colorlinks=true,linkcolor=black,citecolor=black,filecolor=black,urlcolor=black}

\setcellgapes{3pt}

\graphicspath{{figures/}}

\DeclarePairedDelimiter{\paren}{\lparen}{\rparen}
\DeclarePairedDelimiter{\abs}{\lvert}{\rvert}
\DeclarePairedDelimiter{\ave}{\langle}{\rangle}
\DeclarePairedDelimiter{\braces}{\lbrace}{\rbrace}
\newcommand{\parenpow}[3]{\paren[#1]{#2}^{\!#3}}

\newcommand{\dd}{\mathrm{d}}
\newcommand{\pp}{\partial}
\newcommand{\deriv}[2]{\frac{\dd #1}{\dd #2}}

\DeclareMathOperator{\expm}{expm1}
\DeclareMathOperator{\sgn}{sgn}

\newcommand{\yy}{y_m}
\newcommand{\yinit}{y^\text{(init)}_m}
\newcommand{\yfive}{y^\text{(5)}_m}
\newcommand{\yfour}{y^\text{(4)}_m}
\newcommand{\fabs}{f_\mathrm{abs}}
\newcommand{\frel}{f_\mathrm{rel}}

\newcommand{\s}{\mathrm{s}}
\newcommand{\cm}{\mathrm{cm}}
\newcommand{\nm}{\mathrm{nm}}
\newcommand{\g}{\mathrm{g}}
\newcommand{\hz}{\mathrm{Hz}}
\newcommand{\ghz}{\mathrm{GHz}}
\newcommand{\erg}{\mathrm{erg}}
\newcommand{\sr}{\mathrm{sr}}
\newcommand{\jy}{\mathrm{Jy}}

\newcommand{\mmp}{m_\mathrm{p}}
\newcommand{\mme}{m_\mathrm{e}}
\newcommand{\msun}{M_\odot}
\newcommand{\kb}{k_\mathrm{B}}

\newcommand{\xpix}{x_\mathrm{pix}}
\newcommand{\kpix}{k_\mathrm{pix}}
\newcommand{\kkpix}{k^\mathrm{pix}}
\newcommand{\fstep}{f_\mathrm{step}}
\newcommand{\rhor}{r_\mathrm{hor}}
\newcommand{\dtau}{\Delta\tau}
\newcommand{\dtaumax}{\Delta\tau_\mathrm{max}}
\newcommand{\kappae}{\kappa_\mathrm{e}}
\newcommand{\se}{s_\mathrm{e}}
\newcommand{\nne}{n_\mathrm{e}}
\newcommand{\nni}{n_\mathrm{i}}
\newcommand{\thetae}{\Theta_\mathrm{e}}
\newcommand{\te}{T_\mathrm{e}}
\newcommand{\ti}{T_\mathrm{i}}
\newcommand{\gammamin}{\gamma_\mathrm{min}}
\newcommand{\gammamax}{\gamma_\mathrm{max}}
\newcommand{\rhigh}{R_\mathrm{high}}
\newcommand{\rlow}{R_\mathrm{low}}
\newcommand{\usource}{u_\mathrm{source}}
\newcommand{\ucamera}{u_\mathrm{camera}}

\newcommand{\code}[1]{\texttt{#1}}

\shorttitle{Blacklight}
\shortauthors{C.~J.~White}

\begin{document}

\title{Blacklight:\ A General-Relativistic Ray-Tracing and Analysis Tool}
\author{Christopher~J.~White}
\affiliation{Department of Astrophysical Sciences, Princeton University, Peyton Hall, Princeton, NJ, USA}

\begin{abstract}

We describe the \code{Blacklight} code, intended for post-processing general-relativistic magnetohydrodynamic simulation data. Beyond polarized ray tracing of synchrotron radiation, it can produce a number of outputs that aid in analyzing data sets, such as maps of auxiliary quantities and false-color renderings. Additional features include support for adaptive mesh refinement input, slow-light calculations, and adaptive ray tracing. The code is written with ease of use, readability, and transparency as primary objectives, while it still achieves high performance. \code{Blacklight} is publicly available and released into the public domain.

\strut

\end{abstract}

\section{Introduction}
\label{sec:introduction}

Direct simulation of accreting black holes, employing general-relativistic magnetohydrodynamic (GRMHD) codes, is an indispensable method for understanding these physical systems. However, there is a disconnect separating the directly modeled quantities describing the plasma from the directly observed light at various frequencies. It is fortunate, then, that for supermassive black holes accreting well below the Eddington rate, the observed light in the millimeter-to-infrared range is largely due to optically thin synchrotron radiation emitted by the electrons. Such radiation, energetically unimportant for and neglected by the GRMHD evolution, can be recovered by post-processing simulation data with ray-tracing codes, producing predictions for light curves, spectra, and even resolved images based on such simulations.

Ray tracing in this manner is critical for interpreting Event Horizon Telescope (EHT) observations of the M87 black hole \citep{EHT2019e,EHT2019f,EHT2021} and GRAVITY and EHT observations of Sagittarius~A* (Sgr~A*) \citep{Gravity2018,EHT2022e}, for example, and can be a key ingredient in data-analysis pipelines \citep{Wong2022}. Given that there are a number of GRMHD codes in active use, and that different analyses will make distinct demands upon ray tracing, there are a number of existing ray-tracing codes. These include \code{BHOSS} \citep{Younsi2020}, \code{GRay} \citep{Chan2013} and \code{GRay2} \citep{Chan2018}, \code{grtrans} \citep{Dexter2009,Dexter2016}, \code{ibothros} \citep{Noble2007} and \code{ipole} \citep{Moscibrodzka2018}, \code{Odyssey} \citep{Pu2016,Pu2018}, \code{RAIKOU} \citep{Kawashima2021}, \code{RAPTOR} \citep{Bronzwaer2018,Bronzwaer2020}, and \code{VRT2} \citep[cf.][]{Broderick2003,Broderick2004}, as summarized and compared in \citet{Gold2020}.\footnote{An even earlier application of ray tracing to GRMHD simulations can be found in \citet{Schnittman2006}, though with an emphasis on X-ray emission and stellar-mass black holes.}

A number of these codes are publicly available, though each is adapted to particular GRMHD codes, and none currently natively support \code{Athena++} \citep{Stone2020,White2016} simulation data. Furthermore, the exclusive focus of most codes has been, understandably, producing simple synchrotron images from simulation data,\footnote{Notable exceptions include the interactive modes of \code{GRay}, \code{GRay2}, and \code{Odyssey}, and the virtual-reality data products from \code{RAPTOR} \citep{Davelaar2018}. Interestingly, of the existing codes mentioned, these are precisely the four that work with GPUs, leveraging the ability to rapidly produce a single image to enable these techniques.} while the growing diversity and complexity of the research performed by this community suggest the time has come to consider more general tool sets for analyzing simulations.

Here we describe \code{Blacklight},\footnote{See \url{https://doi.org/10.5281/zenodo.6591886} for all versions of the code and \url{https://doi.org/10.5281/zenodo.6591887} for version 1.0 released concurrently with this work.} a publicly available\footnote{The code is placed into the public domain. It is currently hosted at \url{https://github.com/c-white/blacklight}.} code for state-of-the-art ray-tracing techniques, as well as additional analysis tools related to calculating geodesics in curved spacetimes. \code{Blacklight} is designed with the goal of enabling scientific research; beyond accuracy and performance, primary objectives influencing how the code is written include ease of use (no dependencies on external libraries), readability (clean, consistent coding, using encapsulation when and only when appropriate given the physics and equations being considered), and transparency (thorough documentation both in the source code and in the related wiki\footnote{Currently hosted at \url{https://github.com/c-white/blacklight/wiki}.}). It enables, for the first time, direct usage of \code{Athena++} outputs, including data with static and adaptive mesh refinement. It also supports files generated by the \code{iharm3D} \citep{Prather2021} version of \code{HARM} \citep{Gammie2003}, as well as a legacy \code{HARM} format.

Section~\ref{sec:integration} provides the details for the numerical choices made by \code{Blacklight}, including the integration of geodesics (\S\ref{sec:integration:geodesics}), unpolarized radiation (\S\ref{sec:integration:unpolarized}), and polarized radiation (\S\ref{sec:integration:polarized}). We illustrate the correctness of the code with test problems from the literature in Section~\ref{sec:tests}. Particularly noteworthy capabilities are discussed in Section~\ref{sec:features}, including slow light (\S\ref{sec:features:slow}), adaptive ray tracing (\S\ref{sec:features:adaptive}), integration of quantities other than the intensity of light (\S\ref{sec:features:auxiliary}), false-color renderings (\S\ref{sec:features:renderings}), and the production of ``true-color'' images (\S\ref{sec:features:true}). Performance data can be found in Section~\ref{sec:performance}. We describe the coding considerations and philosophy that characterize \code{Blacklight} in Section~\ref{sec:philosophy}.

We use the $(-,\, +,\, +,\, +)$ metric signature. In purely geometrical contexts (i.e., the metric and null geodesics), we omit factors of $c$ and $G$, though we keep the black hole mass $M$ as a convenient dimensional-consistency check, so that mass, length, time, and spin all have the same, nontrivial unit. When discussing plasma and radiation, $c$ and $G$ are restored, and equations hold in CGS units.

\section{Integration Algorithms}
\label{sec:integration}

\code{Blacklight}, like all such ray tracers, proceeds by defining a camera (essentially an array of pixels), calculating a ray for each pixel in the camera, and calculating the relevant properties of light along each ray. Each pixel defines the initial conditions (spacetime position and momentum) for its associated ray (null geodesic). Rays are calculated by integrating the geodesic equation on a stationary spacetime, starting at the camera and going backward in time to the source. The appropriate radiative transfer equation (for unpolarized light, polarized light, or some auxiliary quantity) is then integrated along each ray from the source to the camera.

\newpage

The calculations are performed in Cartesian Kerr--Schild coordinates\footnote{These coordinates are used by \code{GRay2}, though most existing ray tracers employing a fixed metric choose spherical Kerr--Schild or Boyer--Lindquist coordinates.} $x^\alpha = (t,\, x,\, y,\, z)$ appropriate for a Kerr black hole with mass $M$ and dimensionless spin $a / M$, where the metric components are
\begin{subequations} \begin{align}
  g_{\alpha\beta} & = \eta_{\alpha\beta} + f l_\alpha l_\beta, \\
  g^{\alpha\beta} & = \eta^{\alpha\beta} - f l^\alpha l^\beta.
\end{align} \end{subequations}
Here $\eta$ is the Minkowski metric, and we define
\begin{subequations} \label{eq:cks_aux} \begin{align}
  f & = \frac{2 M r^3}{r^4 + a^2 z^2}, \\
  l_\alpha,\, l^\alpha & = \braces[\bigg]{\pm1,\, \frac{r x + a y}{r^2 + a^2},\, \frac{r y - a x}{r^2 + a^2},\, \frac{z}{r}}, \\
  2 r^2 & = R^2 - a^2 + \parenpow{\Big}{\parenpow{\big}{R^2 - a^2}{2} + 4 a^2 z^2}{1/2}, \label{eq:cks_aux:r} \\
  R^2 & = x^2 + y^2 + z^2. \label{eq:cks_aux:rr}
\end{align} \end{subequations}
These coordinates are horizon penetrating, though null geodesics traced backward can only asymptotically approach the horizon with infinite affine parameter. More importantly, they have no polar axis singularity, which might otherwise make accurate integration difficult in its vicinity. We will index these coordinates with $\alpha$, $\beta$, $\gamma$, and $\delta$ (spacetime) or $a$ and $b$ (space).

At times we will make reference to spherical Kerr--Schild coordinates $x^\mu = (t,\, r,\, \theta,\, \phi)$, indexed with $\mu$ and $\nu$ (spacetime) or $i$ and $j$ (space). These are related to their Cartesian counterparts via \eqref{eq:cks_aux:r} and \eqref{eq:cks_aux:rr}, together with
\begin{subequations} \begin{align}
  \theta & = \cos^{-1}\paren[\bigg]{\frac{z}{r}}, \\
  \phi & = \tan^{-1}(y, x) - \tan^{-1}\paren[\bigg]{\frac{a}{r}}.
\end{align} \end{subequations}
The inverse transformation is
\begin{subequations} \begin{align}
  x & = \sin\theta \, (r \cos\phi - a \sin\phi), \\
  y & = \sin\theta \, (r \sin\phi + a \cos\phi), \\
  z & = r \cos\theta.
\end{align} \end{subequations}
For completeness, we note the metric components are
\begin{widetext} \begin{equation}
  g_{\mu\nu} =
  \begin{pmatrix}
    -(1 - f) & f & 0 & -a f s^2 \\
    f & 1 + f & 0 & -a (1 + f) s^2 \\
    0 & 0 & 2 M r / f & 0 \\
    -a f s^2 & -a (1 + f) s^2 & 0 & (r^2 + a^2 + a^2 f s^2) s^2
  \end{pmatrix},
\end{equation} \end{widetext}
where $s = \sin\theta$.

We will also make reference to the normal frame related to spherical Kerr--Schild coordinates, indexing this with $n$ for time and $p$ and $q$ for space. Vectors transform between the two according to
\begin{subequations} \begin{align}
  A^n & = \alpha A^t, \\
  A^p & = \delta^p_i \paren[\bigg]{A^i + \beta^i A^t},
\end{align} \end{subequations}
with $\alpha = (-g^{tt})^{-1/2}$ the lapse and $\beta^i = -g^{ti} / g^{tt}$ the shift. This frame, familiar to those working with $3{+}1$ decompositions, has metric components $g_{nn} = -1$, $g_{np} = 0$, and $g_{pq} = \delta^i_p \delta^j_q g_{ij}$.

Blacklight can use simulation data in either Cartesian or spherical Kerr--Schild coordinates, transforming values in the latter case into a singularity-free system for processing.

\subsection{Camera Definition}
\label{sec:integration:camera}

There are two common types of camera used in general-relativistic ray tracing, which we will term ``plane-parallel'' and ``pinhole.'' The former takes pixels to be located at distinct points in a plane (suitably defined), each seeing light with the same parallel (again, suitably defined) momentum. The latter takes all pixels to be collocated in space but sensitive to different momentum directions. Plane-parallel cameras are appropriate for making images as seen at infinity without having to place the camera at large distances from the source, while pinhole cameras model what a small detector would see at its location. Codes in the literature use one or the other, and \code{Blacklight} supports both.

In either case, \code{Blacklight} requires the user to specify the position $x^i$ of the camera center, the normal-frame velocity $u^p$ of the camera center, and the unnormalized momentum $k_i = \delta_i^p k_p$ of the light received at the camera center. Velocities are specified as $u^p$ rather than $u^i$, as the latter fails to uniquely determine a future-directed $4$-velocity within the ergosphere.

With the components of $x$, $u$, and $k$ known in any coordinate system, we can define unit vectors for the line-of-sight, vertical, and horizontal directions. To do this we work in the camera frame based on Cartesian Kerr--Schild coordinates, where the metric has components
\begin{subequations} \begin{align}
  g_{t'a'} & = -\delta_{ta}, \\
  g_{a'b'} & = g_{ab} - \frac{u_a}{u_t} g_{tb} - \frac{u_b}{u_t} g_{ta} + \frac{u_a u_b}{u_t u_t} g_{tt}, \\
  g^{t'a'} & = -\delta^{ta}, \\
  g^{a'b'} & = g^{ab} + u^a u^b.
\end{align} \end{subequations}

The line-of-sight vector $K$ is a unit vector parallel to $k$. Explicitly, we have
\begin{subequations} \begin{align}
  K_{a'} & \propto k_a - \frac{u_a}{u_t} k_t, \\
  K^{t'} & \propto -u^\alpha k_\alpha, \\
  K^{a'} & \propto g^{a'b'} K_{b'},
\end{align} \end{subequations}
where the proportionality constant is set by requiring $K_{a'} K^{a'} = 1$. The transformation rule
\begin{subequations} \begin{align}
  K^t & = u^t K^{t'} - \frac{u_a}{u_t} K^{a'}, \\
  K^a & = K^{a'} + u^a K^{t'}
\end{align} \end{subequations}
returns these components to the coordinate frame.

A vertical direction $v$ begins with a fiducial ``up'' vector $U^{a'} = \delta^a_z$. In the special case that the camera is centered on the polar ($z$) axis, we instead take $U^{a'} = \delta^a_y$. Then $v$ is defined from $U$ orthogonal to $K$ via Gram--Schmidt:
\begin{subequations} \begin{align}
  v^{t'} & \propto 0, \\
  v^{a'} & \propto U^{a'} - U^{b'} K_{b'} K^{a'}, \\
  v_{a'} & \propto g_{a'b'} v^{b'},
\end{align} \end{subequations}
where the normalization is determined by $v_{a'} v^{a'} = 1$. The orthogonal horizontal direction is defined as
\begin{subequations} \begin{align}
  h^{t'} & = 0, \\
  h^{a'} & = \frac{1}{\sqrt{D}} [a\ b\ c] v_{b'} K_{c'},
\end{align} \end{subequations}
where $D$ is the determinant of $g_{a'b'}$ and $[a\ b\ c]$ is the Levi--Civita symbol. \code{Blacklight} allows users to rotate the camera about its axis by an angle $\psi$, so that the horizontal and vertical directions become
\begin{subequations} \begin{align}
  H & = h \cos\psi - v \sin\psi, \\
  V & = v \cos\psi + h \sin\psi.
\end{align} \end{subequations}

Users specify a camera width $w$ and linear resolution $N$. Consider a pixel with horizontal and vertical indices $i,\, j \in \{0,\, \ldots,\, N - 1\}$. Define the coefficients
\begin{equation}
  a,\, b = \paren[\bigg]{\{i,\, j\} - \frac{N}{2} + \frac{1}{2}} \frac{w}{N}.
\end{equation}
For plane-parallel cameras, the pixel's position and momentum are defined to be
\begin{subequations} \begin{align}
  \xpix^\alpha & = x^\alpha + \Delta x^\alpha, \\
  \kpix^a & = K^a,
\end{align} \end{subequations}
where the displacement in the camera frame is
\begin{equation}
  \Delta x^{\alpha'} = a H^{\alpha'} + b V^{\alpha'}.
\end{equation}
For pinhole cameras, they are
\begin{subequations} \begin{align}
  \xpix^\alpha & = x^\alpha, \\
  \kpix^a & = \kpix^{a'} + u^a \kpix^{t'},
\end{align} \end{subequations}
where we have
\begin{subequations} \begin{align}
  \kpix^{t'} & = K^t, \\
  \kpix^{a'} & = \frac{r K^a - a H^a - b V^a}{(r^2 + a^2 + b^2)^{1/2}}.
\end{align} \end{subequations}

For both cameras, the remaining component $\kpix^t$ is found such that $g_{\alpha\beta} \kpix^\alpha \kpix^\beta = 0$. There is a unique positive root on and outside the ergosphere. Inside the ergosphere, the lesser of the two positive roots is chosen. Note the normalization of $\kpix$ is done in an arbitrary but well-defined way here without reference to the frequency of light being observed, fixing the scale of the affine parameter used in the geodesic equation.

Eventually, the physical frequency $\nu$ (e.g., $230\ \ghz$) must come into play. At the camera, the ray's physical momentum will be $n \kpix$, where the normalization $n$ is set by one of two user options. If the camera is to record frequency $\nu$ at its location (most appropriate for pinhole cameras), then
\begin{equation}
  -n \kkpix_\alpha u^\alpha = \nu.
\end{equation}
Alternatively, if the camera is to record light that would be observed at frequency $\nu$ by a static observer at infinity (most appropriate for plane-parallel cameras), then
\begin{equation}
  -n \kkpix_t = \nu.
\end{equation}

\subsection{Geodesics}
\label{sec:integration:geodesics}

Given initial contravariant position components $x^\alpha$ and covariant null momentum components $k_\alpha$, \code{Blacklight} integrates the geodesic equation in the Hamiltonian form
\begin{subequations} \label{eq:geodesic} \begin{align}
  \deriv{x^\alpha}{\lambda} & = g^{\alpha\beta} k_\beta, \\
  \deriv{k_t}{\lambda} & = 0, \\
  \deriv{k_a}{\lambda} & = -\frac{1}{2} \pp_a g^{\alpha\beta} k_\alpha k_\beta.
\end{align} \end{subequations}
\citep[cf.][]{James2015,Pu2016,Kawashima2021,VelasquezCadavid2022}. It also sometimes tracks proper distance along the ray via
\begin{equation} \label{eq:length}
  \deriv{s}{\lambda} = \parenpow{\big}{g_{pq} k^p k^q}{1/2}.
\end{equation}

Integration proceeds in the direction of negative $\lambda$ (backward in time) from the camera. It terminates when the radial coordinate $r$ crosses the threshold from less than to greater than that of the camera center. Even in horizon-penetrating coordinates, backward-propagating rays cannot reach the horizon in finite affine parameter. Geodesics are therefore also terminated if they cross inside an inner $r$ threshold set by the user.

\code{Blacklight} offers three numerical schemes for integrating \eqref{eq:geodesic}. One is the standard second-order Runge--Kutta (RK2) Heun's method, which has the Butcher tableau given in Table~\ref{tab:rk2}, and another is the standard fourth-order Runge--Kutta (RK4) method with the Butcher tableau of Table~\ref{tab:rk4}. In both cases, step size in affine parameter is taken to be
\begin{equation} \label{eq:rk_step}
  \Delta\lambda = \fstep (r - \rhor),
\end{equation}
where $\fstep$ is a user-specified constant and $\rhor$ is the outer horizon radius.

\begin{table}
  \caption{Butcher tableau for the RK2 geodesic integrator. \label{tab:rk2}}
  \begin{equation*}
    \begin{array}{c|cc}
      0 \\
      1 & 1 \\
      \hline
      & 1/2 & 1/2
    \end{array}
  \end{equation*}
\end{table}

\begin{table}
  \caption{Butcher tableau for the RK4 geodesic integrator. \label{tab:rk4}}
  \begin{equation*}
    \begin{array}{c|cccc}
      0 \\
      1/2 & 1/2 \\
      1/2 & 0 & 1/2 \\
      1 & 0 & 0 & 1 \\
      \hline
      & 1/6 & 1/3 & 1/3 & 1/6
    \end{array}
  \end{equation*}
\end{table}

In terms of robustness and ease of use, however, the recommended geodesic integrator is the ``RK5(4)7M'' method of \citet[DP]{Dormand1980}, which is an adaptive fifth-order Runge--Kutta method that calculates fourth-order Runge--Kutta values from the same set of function evaluations in order to estimate errors for each step. In comparing common integrators in the context of the computation of null geodesics around black holes, \citet{VelasquezCadavid2022} find DP to be far more accurate. The step size is adjusted dynamically based on this error estimate, in order to take steps as large as possible given a user-prescribed tolerance \citep[cf.][]{NR}. Here the error is defined in terms of the initial, fourth-order, and fifth-order values of the eight dependent variables $\{\yy\} = \{x^\alpha\} \cup \{k_\alpha\}$ in \eqref{eq:geodesic} via
\begin{equation}
  \epsilon = \max_m\braces[\Bigg]{\frac{\abs[\big]{\yfive - \yfour}}{\fabs + \frel \max\paren[\big]{\abs[\big]{\yinit}, \abs[\big]{\yfive}}}},
\end{equation}
where $\fabs$ and $\frel$ are provided by the user. Values of $\epsilon > 1$ indicate a step must be reattempted with a smaller $\Delta\lambda$.

As described in \citet{Shampine1986}, the function evaluations can be arranged in a different way to provide a fourth-order accurate estimate of the midpoint of a step. \code{Blacklight} takes this midpoint to be the position and momentum of the step for the purposes of radiative transfer integration. In cases where the geodesic steps cover more proper distance than a user-specified limit (such that the radiative transfer equation would be poorly sampled even where the geodesic is calculated accurately), \code{Blacklight} subdivides the step. The aforementioned midpoint, together with function values and derivatives at both endpoints of the step, can be used to uniquely define a quartic expression, providing a fourth-order accurate interpolation anywhere within the step \citep{Shampine1986} and enabling this subdivision. The exact choice of coefficients used in DP implementations varies; the ones used here are given in Table~\ref{tab:dp}.

\begin{table*}
  \makegapedcells
  \caption{Butcher tableau for the DP geodesic integrator. \label{tab:dp}}
  \begin{equation*}
    \begin{array}{c|ccccccc}
      0 \\
      \frac{1}{5} & \frac{1}{5} \\
      \frac{3}{10} & \frac{3}{40} & $\phs$\frac{9}{40} \\
      \frac{4}{5} & \frac{44}{45} & -\frac{56}{15} & \frac{32}{9} \\
      \frac{8}{9} & \frac{19\,372}{6\,561} & -\frac{25\,360}{2\,187} & \frac{64\,448}{6\,561} & -\frac{212}{729} \\
      1 & \frac{9\,017}{3\,168} & -\frac{355}{33} & \frac{46\,732}{5\,247} & $\phs$\frac{49}{176} & -\frac{5\,103}{18\,656} \\
      1 & \frac{35}{384} & $\phs$0 & \frac{500}{1\,113} & $\phs$\frac{125}{192} & -\frac{2\,187}{6\,784} & $\phs$\frac{11}{84} \\
      \hline
      \text{5th-order step} & \frac{35}{384} & $\phs$0 & \frac{500}{1\,113} & $\phs$\frac{125}{192} & -\frac{2\,187}{6\,784} & $\phs$\frac{11}{84} & 0 \\
      \text{4th-order step} & \frac{5\,179}{57\,600} & $\phs$0 & \frac{7\,571}{16\,695} & $\phs$\frac{393}{640} & -\frac{92\,097}{339\,200} & $\phs$\frac{187}{2\,100} & \frac{1}{40} \\
      \text{4th-order midpoint} & \frac{6\,025\,192\,743}{30\,085\,553\,152} & $\phs$0 & \frac{51\,252\,292\,925}{65\,400\,821\,598} & -\frac{2\,691\,868\,925}{45\,128\,329\,728} & $\phs$\frac{187\,940\,372\,067}{1\,594\,534\,317\,056} & -\frac{1\,776\,094\,331}{19\,743\,644\,256} & \frac{11\,237\,099}{235\,043\,384}
    \end{array}
  \end{equation*}
\end{table*}

In all cases, integration is performed in Cartesian Kerr--Schild coordinates. As these coordinates are stationary, use of \eqref{eq:geodesic} ensures the covariant energy is conserved along rays exactly in the code, as it is physically. Because \code{Blacklight} does not employ spherical coordinates here, exact conservation of $\phi$ covariant momentum ($z$ angular momentum) is not guaranteed. However, this choice avoids numerical issues arising from geodesics approaching coordinate poles.

\subsection{Unpolarized Radiation}
\label{sec:integration:unpolarized}

\code{Blacklight} can create unpolarized images, ignoring polarization effects along each ray. In terms of the Lorentz invariants
\begin{subequations} \begin{align}
  I & = I_\nu \nu^{-3}, \\
  j_I & = j_\nu \nu^{-2}, \\
  \alpha_I & = \alpha_\nu \nu,
\end{align} \end{subequations}
the general-relativistic equation for unpolarized radiative transfer is
\begin{equation} \label{eq:unpolarized}
  \deriv{I}{\lambda} = j_I - \alpha_I I,
\end{equation}
where the affine parameter $\lambda$ must be normalized such that \eqref{eq:geodesic} holds. We take $j_I$ and $\alpha_I$ to be constant over each ray segment (typically no larger than nearby cells in the underlying simulation). The evolution of $I$ from the beginning of a segment (where we denote it $I^-$) to the end (where we denote it $I^+$), across an optical depth $\dtau = \alpha_I \Delta\lambda$, employs the well-known exact solution
\begin{widetext} \begin{equation}
  I^+ =
  \begin{cases}
    \exp(-\dtau) \paren[\bigg]{I^- + \expm(\dtau) \dfrac{j_I}{\alpha_I}}, & \alpha_I > 0\ \text{and}\ \dtau < \dtaumax; \\
    \dfrac{j_I}{\alpha_I}, & \alpha_I > 0\ \text{and}\ \dtau \geq \dtaumax; \\
    I^- + j_I \Delta\lambda, & \alpha_I = 0;
  \end{cases}
\end{equation} \end{widetext}
where we use the standard function $\expm(\cdot) = \exp(\cdot) - 1$ as written and fix a finite $\dtaumax = 100$ to avoid floating-point inaccuracies.

\subsection{Polarized Radiation}
\label{sec:integration:polarized}

Users can choose to integrate the equations of polarized radiative transfer, evolving four degrees of freedom and producing images in Stokes parameters $I_\nu$, $Q_\nu$, $U_\nu$, and $V_\nu$ at the camera. In addition to $j_I$ and $\alpha_I$, this requires knowing the Lorentz-invariant linearly polarized emissivities $j_Q$ and $j_U$, circularly polarized emissivity $j_V$, linearly polarized absorptivities $\alpha_Q$ and $\alpha_U$, circularly polarized absorptivity $\alpha_V$, Faraday conversion rotativities $\rho_Q$ and $\rho_U$, and Faraday rotation rotativity $\rho_V$.

There are two primary approaches used by polarized general-relativistic ray tracing codes. Some (e.g., \code{grtrans}) make use of the convenient form of the Walker--Penrose constant \citep{Walker1970} along null geodesics in Boyer--Lindquist coordinates to parallel-transport fiducial polarization directions backward from the camera, thus allowing each segment's local definition of the Stokes parameters to match those at the camera. Here we adopt the alternative approach of evolving the polarized state of the radiation forward in a way commensurate with both plasma and coordinate effects, as done in, e.g., \code{ipole}. This allows our use of Cartesian (or even more general) coordinates.

The method we use is well described by \citet{Moscibrodzka2018}; we only briefly summarize it here. The polarized transfer equation can be written as both
\begin{equation} \label{eq:polarized_s}
  \deriv{S_A}{\lambda} = \underbrace{J_A - M_{AB} S_B}_\text{plasma terms}\ +\ \text{(coordinate terms)}
\end{equation}
and
\begin{equation} \label{eq:polarized_n}
  \deriv{N^{\alpha\beta}}{\lambda} = \underbrace{-k^\gamma \paren[\big]{\Gamma^\alpha_{\gamma\delta} N^{\delta\beta} - \Gamma^\beta_{\gamma\delta} N^{\alpha\delta}}}_\text{coordinate terms}\ +\ \text{(plasma terms)}.
\end{equation}
Here $A$ and $B$ run over the Stokes indices (with repetition implying summation), $S$ is the vector of Stokes parameters, $J$ is the vector of emissivities, $M$ is the matrix of absorptivities and rotativities, $N$ is the complex Hermitian coherency tensor, and $\Gamma$ is the collection of connection coefficients. The dependent variables are encoded in both $S$ and $N$, either of which can be used to define the other in a given frame.

The general procedure is to use Strang splitting for each ray segment of affine parameter length $\Delta\lambda$, evolving \eqref{eq:polarized_n} without the unwieldy plasma terms for $\Delta\lambda / 2$, then evolving \eqref{eq:polarized_s} without the equally unwieldy coordinate terms for $\Delta\lambda$, and finally evolving \eqref{eq:polarized_n} without plasma coupling for another $\Delta\lambda / 2$. All coefficients are held constant over each substep. As the last substep in one segment and the first substep in the next segment evolve the same equation, we merge these two via a second-order predictor-corrector method with the Butcher tableau given in Table~\ref{tab:pc}.

\begin{table}
  \caption{Butcher tableau for the predictor-corrector evolution of \eqref{eq:polarized_n}. \label{tab:pc}}
  \begin{equation*}
    \begin{array}{c|cc}
      0 \\
      1/2 & 1/2 \\
      \hline
      & 0 & 1
    \end{array}
  \end{equation*}
\end{table}

Just as \eqref{eq:unpolarized} requires special care and exact solutions to avoid numerically unstable integration, so too does \eqref{eq:polarized_s}. The full solution is described in a number of sources, such as \citet{Jones1977}, \citet{LandiDeglInnocenti1985}, and \citet{Moscibrodzka2018}.\footnote{We note there is a sign error in the third of five terms in (24) in \citet{Moscibrodzka2018}; the error is purely typographical and does not appear in the source code for \code{ipole}.} The latter provides useful simplifying cases for when various coefficients vanish, which \code{Blacklight} employs when applicable.

\subsection{Sampling}
\label{sec:integration:sampling}

Given a simulation grid, possibly including mesh refinement, and a set of rays with sample points, \code{Blacklight} interpolates values from the former onto the latter. This interpolation is done in the primitive variables employed by \code{Athena++}:\ total rest-mass density $\rho$; total pressure $p$ or electron entropy $\kappae$, depending on electron model (see \S\ref{sec:integration:coefficients}); fluid velocity $u^p$, or the Cartesian Kerr--Schild normal-frame values if appropriate; and magnetic field $B^i$ or $B^a$, as appropriate.

Interpolation can be done in nearest-neighbor fashion, where the values are copied from whatever unique cell bounds the sample point. Alternatively, \code{Blacklight} can perform trilinear interpolation in simulation coordinates. \code{Athena++} domain-decomposes the simulation into logically Cartesian blocks of cells, and interpolation within a block costs negligibly more than nearest-neighbor lookup. Accurate interpolation across block boundaries, however, can dominate the cost of producing an image, given that adjacent blocks may be at different refinement levels and ghost zones are generally not included in data dumps. Interblock interpolation rarely makes a noticeable difference in the results, compared to intrablock interpolation; users can choose which method to employ.

The plasma state at each sample point is fully defined after interpolation. Useful quantities that can be derived from the interpolated ones, possibly with user-defined assumptions about the plasma, include electron number density $\nne$; electron dimensionless temperature $\thetae = \kb \te / \mme c^2$, with $\te$ the electron temperature; fluid-frame magnetic field magnitude $B = (b_\alpha b^\alpha)^{1/2}$, with $b^\alpha = u_\beta (*F)^{\beta\alpha}$ the components of the magnetic $4$-vector and $F$ the electromagnetic field tensor; plasma $\sigma = B^2 / \rho$; and plasma $\beta^{-1} = B^2 / 2 p$.

\subsection{Coefficients}
\label{sec:integration:coefficients}

It is standard practice to rotate one's coordinate system at each point along a ray in order to make $j_U$, $\alpha_U$, and $\rho_U$ vanish. For the remaining eight coefficients, \code{Blacklight} uses the fitting formulas provided by \citet{Marszewski2021}, which conveniently unifies and corrects the synchrotron literature. These formulas cover the cases of thermal, power-law, and kappa distributions for the electrons, always assumed to be isotropic. \code{Blacklight} can calculate the coefficients resulting from any admixture of these three populations. Thermal electrons are distributed in Lorentz factor $\gamma$ according to the Maxwell--J\"uttner distribution based on $\thetae$:
\begin{equation}
  \frac{\dd\nne}{\dd\gamma} = \frac{\nne \gamma \sqrt{\gamma^2 - 1}}{\thetae K_2(\thetae^{-1})} \exp\paren[\bigg]{-\frac{\gamma}{\thetae}},
\end{equation}
where $K_2$ is the cylindrical Bessel function. Power-law electrons are defined by an index $p$ and Lorentz-factor cutoffs $\gammamin$ and $\gammamax$:
\begin{equation}
  \frac{\dd\nne}{\dd\gamma} \propto
  \begin{cases}
    \gamma^{-p}, & \gamma \geq \gammamin \text{ and } \gamma \leq \gammamax; \\
    0, & \gamma < \gammamin \text{ or } \gamma > \gammamax.
  \end{cases}
\end{equation}
The kappa distribution is defined by index $\kappa$ and width $w$:
\begin{equation}
  \frac{\dd\nne}{\dd\gamma} \propto \gamma \sqrt{\gamma^2 - 1} \parenpow{\bigg}{1 + \frac{\gamma - 1}{\kappa w}}{-\kappa - 1}.
\end{equation}
In all cases, the electron number density is given by
\begin{equation}
  \nne = \frac{\rho}{\mu \mmp} \parenpow{\bigg}{1 + \frac{1}{\nne / \nni}}{-1},
\end{equation}
where the user specifies the molecular weight $\mu$ and electron-to-ion number-density ratio $\nne / \nni$.

Unlike the power-law and kappa-distribution parameters, $\te$ varies across the simulation domain. As the simulations for which \code{Blacklight} is intended often treat the plasma as a single fluid with a single temperature $T$, we must invoke a prescription for determining $\te$ as a function of the local plasma state available in post-processing. The code supports the common choice of declaring a plasma-beta-dependent ratio between the ion temperature and electron temperature,
\begin{equation}
  \frac{\ti}{\te} = \frac{\rhigh + \beta^{-2} \rlow}{1 + \beta^{-2}}
\end{equation}
with global constants $\rhigh$ and $\rlow$, as introduced by \citet{Moscibrodzka2016}. In this case the electron temperature is
\begin{equation}
  \te = \frac{\mu \mmp p}{\kb \rho} \cdot \frac{\nne / \nni + 1}{\nne / \nni + \ti / \te}.
\end{equation}

Alternatively, \code{Blacklight} can utilize the electron entropy available in two-temperature simulations as prescribed by \citet{Ressler2015}. In this case, the code must provide the values of an entropy-like variable $\kappae \propto \exp(\se / \kb)$, where $\se$ is the electron entropy per electron. Then the electron temperature is given by
\begin{equation}
  \te = \frac{\mme c^2}{5 \kb} \paren[\Big]{\parenpow{\big}{1 + 25 (\mme \nne \kappae)^{2/3}}{1/2} - 1}
\end{equation}
\citep[cf.][]{Sadowski2017}.

\section{Test Problems}
\label{sec:tests}

The most comprehensive comparison of ray tracing codes to date is given by \citet{Gold2020}. Here we apply to \code{Blacklight} the two quantitative tests from that work used to compare all codes, illustrating the correctness of geodesics and unpolarized images produced from simple plasma models.

\subsection{Null Geodesic Deflection}
\label{sec:tests:geodesics}

For null geodesics confined to the midplane in Kerr spacetime and outside the photon ring(s), solutions to the geodesic equation involve relatively tractable integrals. In \citet{Gold2020}, the codes' values of the azimuthal deflection angles $\Delta\phi$ of such geodesics are compared to the values given by formulas in \citet{Iyer2009} for the asymptotic deflection from past to future infinity. Here we repeat this test with the following modification. Given that the geodesic origin (source) and termination point (camera) are at finite distance ($r = 1000\ G M / c^2$ here), we compare the result of ray tracing to the finite-distance formulas from \citeauthor{Iyer2009},
\begin{widetext} \begin{subequations} \label{eq:deflection} \begin{align}
  \Delta\phi & = \int_{\usource}^{u_0} \abs[\bigg]{\deriv{\phi}{u}} \, \dd u - \int_{u_0}^{\ucamera} \abs[\bigg]{\deriv{\phi}{u}} \, \dd u, \\
  \abs[\bigg]{\deriv{\phi}{u}} & = \frac{1 - 2 M (1 - a / b) u}{1 - 2 M u + a^2 u^2} \parenpow{\Bigg}{2 M \parenpow{\bigg}{1 - \frac{a}{b}}{2} u^3 - \paren[\bigg]{1 - \frac{a^2}{b^2}} u^2 + \frac{1}{b^2}}{-1/2}, \\
  r_0 = \frac{1}{u_0} & = \frac{2 \abs{b}}{\sqrt{3}} \parenpow{\bigg}{1 - \frac{a^2}{b^2}}{1/2} \cos\paren[\Bigg]{\frac{1}{3} \cos^{-1}\paren[\bigg]{-\frac{3 \sqrt{3} M (1 - a / b)^2}{\abs{b} (1 - a^2 / b^2)^{3/2}}}},
\end{align} \end{subequations} \end{widetext}
where $u = 1 / r$, $r_0$ is the radial coordinate of closest approach, and $b = -k_\phi / k_t$ is the impact parameter.

\begin{figure}
  \centering
  \includegraphics{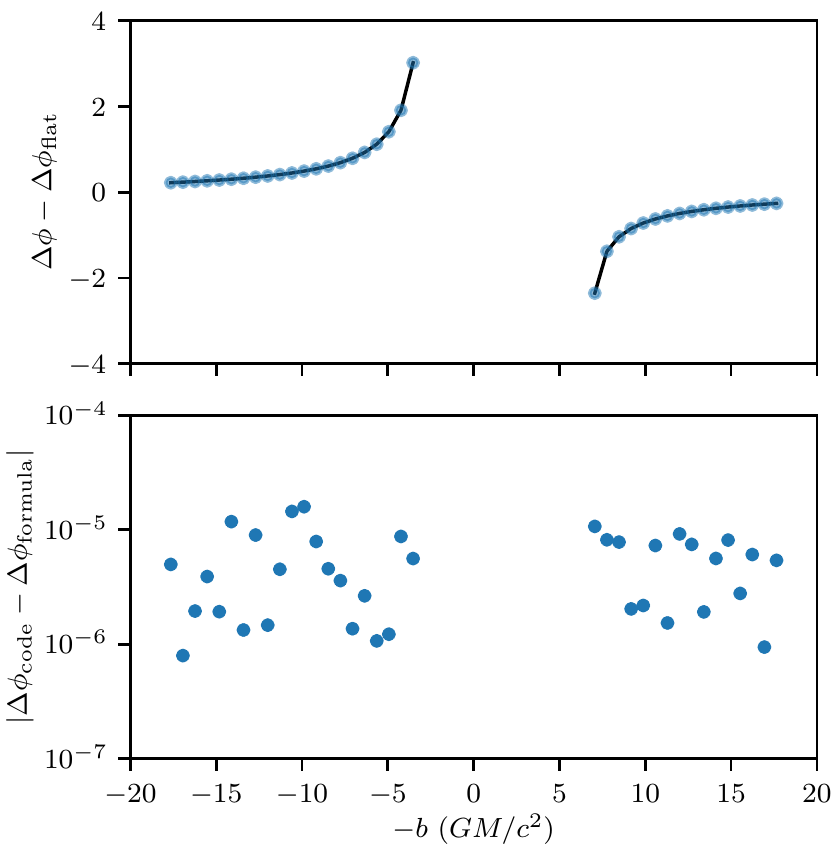}
  \caption{Deflection angles obtained by \code{Blacklight} when integrating the geodesic equation for rays in the midplane around a spinning black hole. The line is calculated from the analytic formula, with the differences shown in the lower panel. This test uses the DP integrator with $\fabs,\, \frel = 10^{-8}$. \label{fig:deflection_values}}
\end{figure}

We place a camera centered at $r = 1000\ G M / c^2$, $\theta = \pi / 2$, and $\phi = 0$ around a black hole with $a / M = 0.9$. There are $51$ pixels across the middle of the image, spanning a width $w = 36\ G M / c^2$. Of these pixels, $14$ are ignored in this analysis; they lie inside the photon ring, given they have impact parameters satisfying $b_- \leq b \leq b_+$ for the critical values
\begin{equation}
  b_\pm = -a \pm 6 M \cos\paren[\Bigg]{\frac{1}{3} \cos^{-1}\paren[\bigg]{\mp \frac{a}{M}}}.
\end{equation}
Figure~\ref{fig:deflection_values} shows the deflection angles calculated by \code{Blacklight} compared to \eqref{eq:deflection}. The left and right sides of the figure correspond to the left and right sides of the image. For plotting purposes, we subtract the asymptotic, flat-spacetime ``deflection'' of $\Delta\phi_\mathrm{flat} = \sgn(b) \pi$. Here we employ the DP integrator with tolerances $\fabs,\, \frel = 10^{-8}$, and the agreement is very good, with absolute errors in the range $10^{-6}\text{--}10^{-5}$. A similar test in \citet[cf.\ their Figure~1]{Gold2020} shows most codes' errors being $10^{-5}$ or greater, though in that case the finite-distance code values are compared to the asymptotic formula.

\begin{figure}
  \centering
  \includegraphics{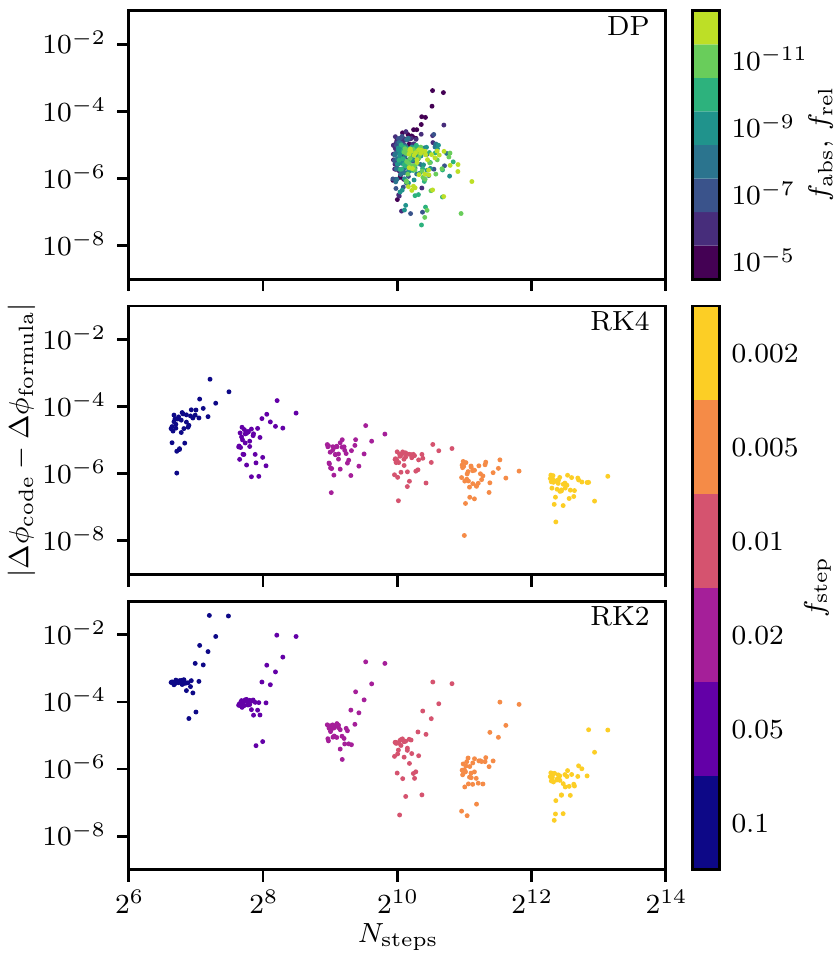}
  \caption{Errors in midplane Kerr deflection angles using the DP, RK4, and RK2 schemes. The error for each ray is plotted against the total number of steps used to integrate that ray. Colors correspond to the user-specified numerical control parameters. The DP integrator is a robust choice, with the RK methods providing versatility for users needing more control over the computation--accuracy trade-off. \label{fig:deflection_errors}}
\end{figure}

We can compare the errors obtained with other numerical tolerances and integration schemes. The top panel of Figure~\ref{fig:deflection_errors} shows the errors for the DP integrator using tolerances ranging over $10^{-12}\text{--}10^{-5}$, plotted against the number of steps taken to trace each ray. The adaptive nature of this integrator means stricter tolerances can be satisfied by only using slight more points near closest approach. For this reason, we recommend DP over the RK integrators for general-purpose ray tracing, where reasonable tolerances set by the user almost always yield reasonably accurate results without the danger of requiring intractably many steps.

When testing the RK4 and RK2 integrators, we vary $\fstep$ over $0.002\text{--}0.1$. The solution improves with decreasing step size, as expected, sometimes even outperforming typical DP results. However, this accuracy can come at a large cost in terms of number of steps taken. Conversely, the RK integrators allow very quick and somewhat inaccurate integrations using relatively few steps. Relative to DP, they permit users to make larger trade-offs between accuracy and performance. It is possible that these non-adaptive variable-step-size methods could be improved for this type of problem with a better heuristic than \eqref{eq:rk_step}.

\subsection{Model Images}
\label{sec:tests:images}

\begin{figure*}
  \centering
  \includegraphics{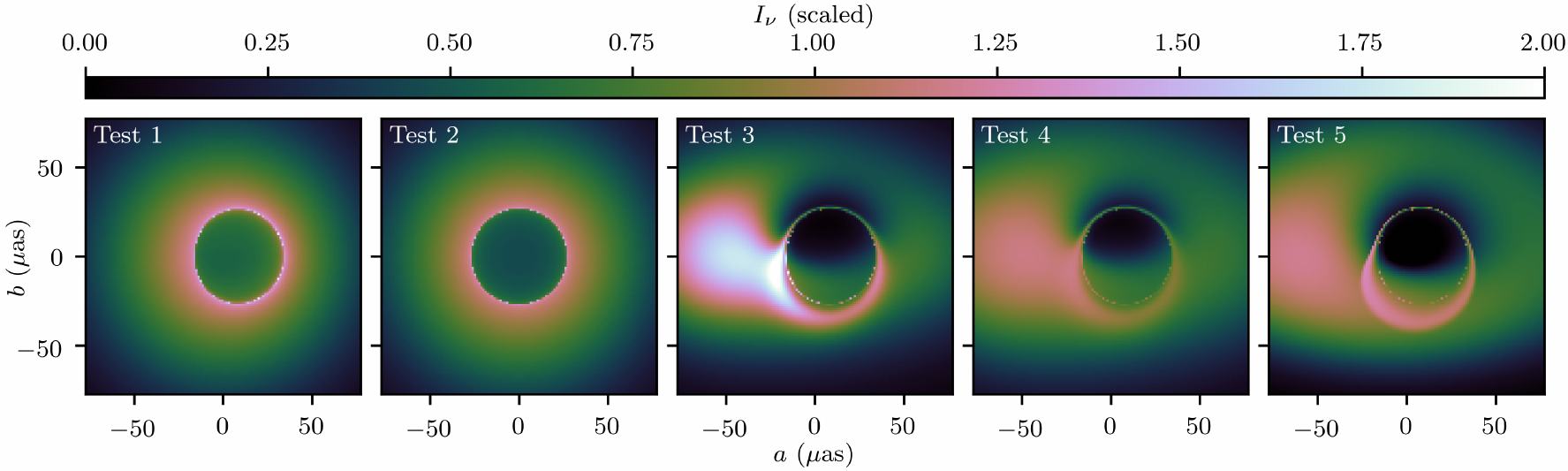}
  \caption{Images produced by \code{Blacklight} for the five parameterized models in \citet{Gold2020}, showing the expected features and brightnesses. In order to match the color scales in that work, $I_\nu$ in $\erg\ \s^{-1}\ \cm^{-2}\ \sr^{-1}\ \hz^{-1}$ is multiplied by $10^5 (75 / 128)^2 f^{-1}$, where $f$ is $1.6465$, $1.436$, $0.4418$, $0.2710$, and $0.0255$ for the five tests, respectively. \label{fig:formula}}
\end{figure*}

\Citet{Gold2020} compare the images produced by unpolarized transfer using five sets of parameterized formulas for velocity, emissivity, and absorptivity, shown in their Figure~2. We repeat these tests with the same parameters and camera settings as in that work, choosing the DP integrator, a plane-parallel camera, and frequencies of $230\ \ghz$ at the location of the camera. The resulting images, shown in Figure~\ref{fig:formula}, agree with those obtained by other codes.

We report the total fluxes for these images in Table~\ref{tab:formula}. Comparing to Table~2 of \citet{Gold2020}, \code{Blacklight} lies comfortably in the middles of the ranges spanned by the seven codes in that work. Our table shows how the minimum and maximum values reported in that comparison deviate from the values obtained here, with the typical minima $1\%$ lower than the \code{Blacklight} value and the typical maxima $1\%$ higher.

\begin{deluxetable}{cCCC}
  \tablecaption{Total fluxes in parameterized model tests. \label{tab:formula}}
  \tablehead{\colhead{} & \colhead{\code{Blacklight}} & \colhead{Code-comparison} & \colhead{Code-comparison} \\[-2ex]
  \colhead{Test} & \colhead{value ($\jy$)} & \colhead{minimum\tablenotemark{a} (\%)} & \colhead{maximum\tablenotemark{a} (\%)}}
  \startdata
  1 & 1.6602 & -0.82 & +0.56 \\
  2 & 1.4494 & -0.92 & +1.49 \\
  3 & 0.4464 & -1.00 & +0.99 \\
  4 & 0.2730 & -0.78 & +1.20 \\
  5 & 0.0257 & -1.22 & +1.12 \\
  \enddata
  \tablenotetext{a}{Deviations of the least and greatest values found by the seven codes in \citet{Gold2020}, relative to the \code{Blacklight} value.}
\end{deluxetable}

\section{Code Features}
\label{sec:features}

\code{Blacklight} has a number of features beyond simply integrating the monochromatic equation of radiative transfer, all of which are intended to aid in the analysis or presentation of simulation data. Some features change how that data is sampled (data cuts, flat spacetime, slow light, and adaptive ray tracing), while others change what quantities are projected onto the image (nonthermal electrons, auxiliary images, false-color renderings, and ``true-color'' images). Each of these features can be toggled independently, working with any others.

In illustrating these features, we employ a fiducial data set and camera parameters. The physical system is that of the thick, aligned disk from \citet{White2020} around a black hole with dimensionless spin $0.9$, with the free scales arbitrarily tuned to match values appropriate for Sgr~A*:\ a black hole mass of $4.152 \times 10^6\ \msun$ \citep{Gravity2019} and an average flux of $2.4\ \jy$ \citep{Doeleman2008}. Regions with plasma $\sigma$ exceeding unity are treated as vacuum.

\begin{figure*}
  \centering
  \includegraphics{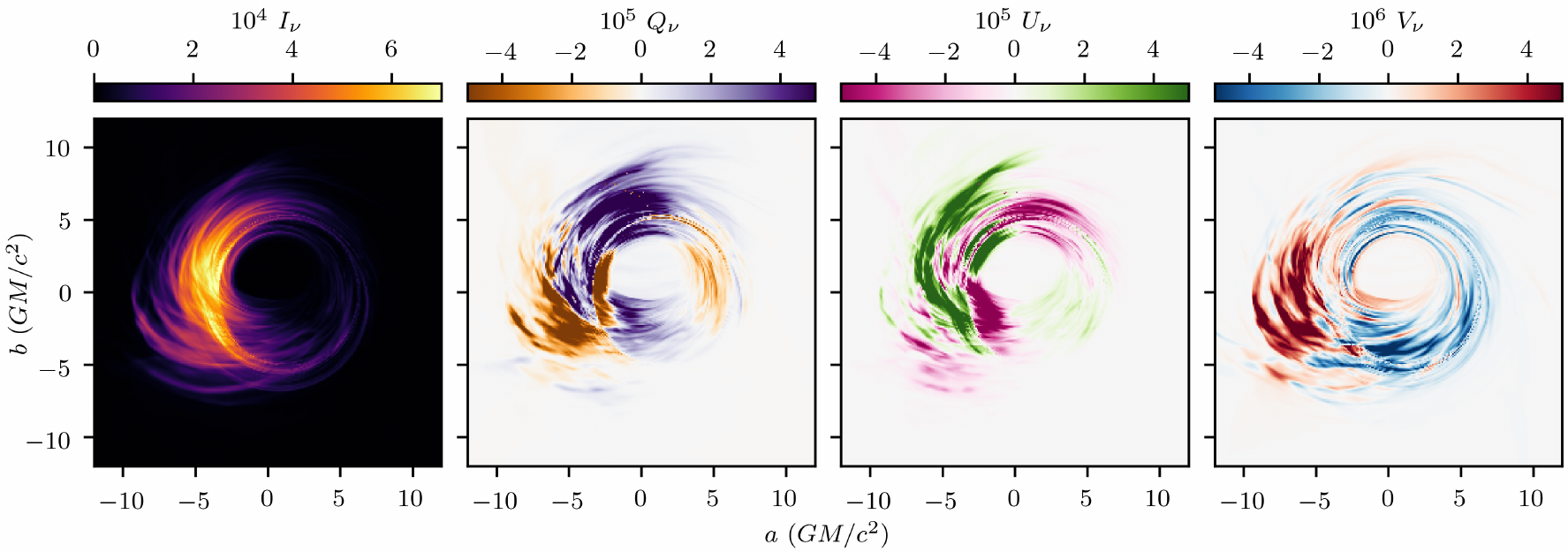}
  \caption{Stokes parameters for the fiducial image, in units of $\erg\ \s^{-1}\ \cm^{-2}\ \sr^{-1}\ \hz^{-1}$. \label{fig:stokes}}
\end{figure*}

The fiducial camera is plane-parallel, with full width $24\ G M / c^2$. Its center is held static at Kerr--Schild $r = 100\ G M / c^2$ and $\theta = 45^\circ$, receiving light with $k_i \propto (1, 0, 0)$. Unless otherwise stated, it collects light that reaches $230\ \ghz$ at infinity, using $256^2$ pixels. Figure~\ref{fig:stokes} shows the four Stokes parameters for the fiducial image.

\newpage

\subsection{Data Cuts}
\label{sec:features:cuts}

It is already common practice to treat any part of a simulation with $\sigma \gtrsim 1$ as vacuum, given that GRMHD jets are often numerically mass loaded and would be unrealistically bright if simulated values there were taken at face value. In fact, this cut is employed in all simulation images shown in this work. Given the importance of probing the origin of image features, \code{Blacklight} offers a simple user interface for making similar cuts on a variety of local conditions. Upper and lower cutoffs can be specified for $\rho$, $\nne$, $p$, $\thetae$, $B$, $\sigma$, and/or plasma $\beta^{-1}$.

\begin{figure}
  \centering
  \includegraphics{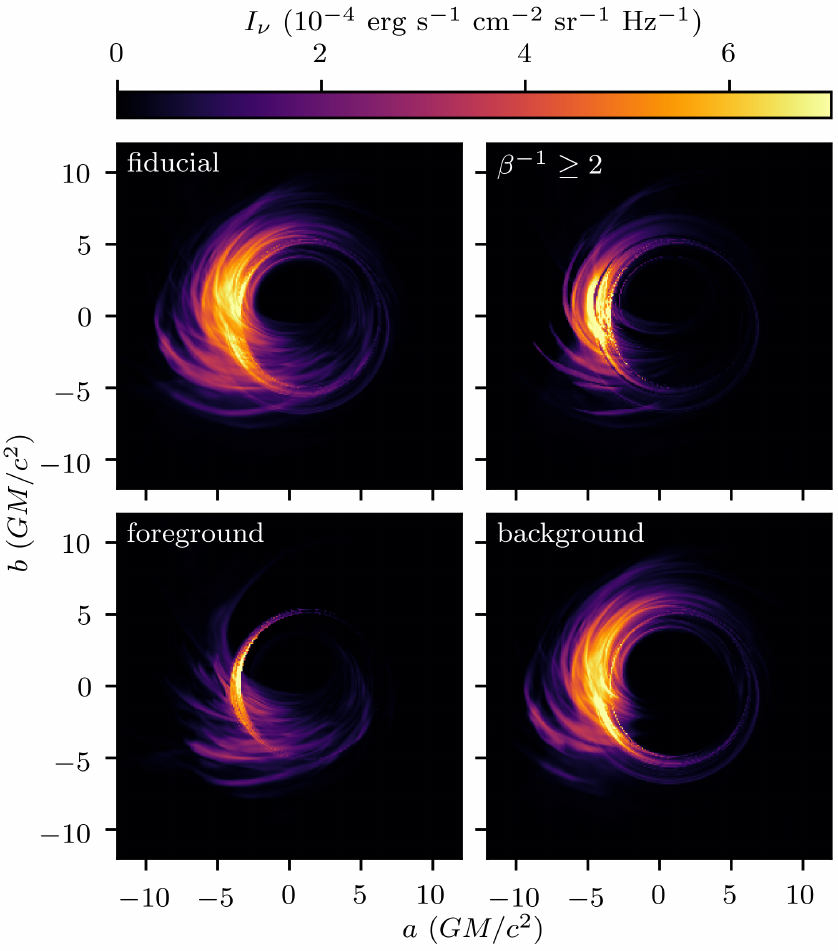}
  \caption{Comparison of the fiducial image (upper left), made with only the standard $\sigma \leq 1$ cut, with the result of various other cuts on the data. Only sufficiently magnetized ($\beta^{-1} \ge 2$) plasma contributes to the upper right image. The lower panels exclude plasma further than (left) or nearer than (right) the plane orthogonal to the line of sight passing through the origin. \label{fig:cuts}}
\end{figure}

As an illustration, the top panels of Figure~\ref{fig:cuts} show the fiducial image and the same image made by considering $\beta^{-1} < 1 / 2$ to be vacuum, highlighting only the coronal regions of the simulation. Some parts of the image become brighter with this cut, indicating the full data set results in some foreground absorption by less magnetized plasma.

\code{Blacklight} also provides users with the ability to cut the data based on position. Users are provided with upper and lower cutoffs on radial coordinate $r$. They can also cut data on one side or another of an arbitrary plane. The plane is defined to be all points $x$ such that $(x^a - x_0^a) n_a = 0$ for a user-provided origin $x_0$ and normal $n$. The lower panels of Figure~\ref{fig:cuts} show what happens when this plane is chosen to bisect the domain into a foreground and a background relative to the camera.

\subsection{Flat Spacetime}
\label{sec:features:flat}

While having gravity bend geodesics with respect to even a Cartesian coordinate system is critical for obtaining accurate images of light emitted by the modeled system, the resulting mapping from the 3D space of a simulation to a 2D image is very non-intuitive. As an analysis tool intended to be more general than a mere imager, \code{Blacklight} allows the user to toggle the effect of gravity. The resulting images more closely correspond to the projections of data one might obtain with standard 3D visualization software, showing features ``where they are'' rather than where their light reaches the camera.

There is ambiguity in how to treat spacetime as flat in this way. \code{Blacklight} uses the following procedure. Simulation data is transformed to Cartesian Kerr--Schild coordinates, both in terms of fluid elements' positions and in terms of the vector components. That is, we take the simulation to provide, perhaps in a roundabout way, $\rho$, $p$ or $\kappae$, $u^\alpha$, and $b^\alpha$, all as functions of $x^\alpha$. These values are then reinterpreted to be functions of Cartesian Minkowski coordinates, with vector components interpreted in that coordinate system. Geodesics are calculated in flat spacetime, and the radiative transfer equation integrates the reinterpreted quantities in flat spacetime as well. Qualitatively similar results could be obtained by, e.g., transforming the simulation to spherical Kerr--Schild and then reinterpreting values as being in spherical Minkowski coordinates.

\begin{figure}
  \centering
  \includegraphics{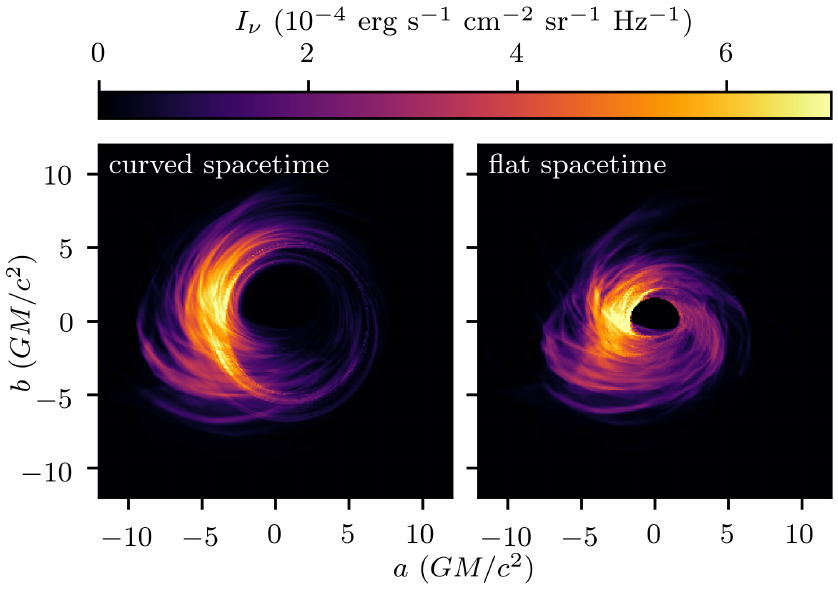}
  \caption{Comparison of the fiducial image (left) with the result of treating spacetime as flat (right). The effect of gravity here is to magnify the image relative to flat-spacetime expectations, as well as to produce a photon ring. \label{fig:flat}}
\end{figure}

The effect on images of artificially assuming a flat spacetime is illustrated in Figure~\ref{fig:flat}. Though this technique does not produce images as would be observed in nature, it can be useful to identify which image features arise as a result of lensing. In the example shown, one can clearly see that the black hole shadow is magnified by gravity. This feature is particularly useful when used in conjunction with false-color renderings (\S\ref{sec:features:renderings} below).

\subsection{Slow Light}
\label{sec:features:slow}

Most ray tracing is calculated with the fast-light approximation, whereby each time slice from a simulation is taken, in turn, to represent the system in a time-invariant state while light propagates from the plasma to the camera. This is sufficient for many purposes, but the emitting region can be tens or hundreds of gravitational times across, while the dynamical times can be less than ten gravitational times, with turbulence and caustic-crossing timescales even shorter. Thus certain applications, especially those concerned with high-frequency variability, require slow-light ray tracing algorithms that account for the time-evolution of the system.

\code{Blacklight} can read a sequence of simulation outputs, sampling the entire spacetime volume needed to make an image. Whenever a ray needs simulation values from a given spacetime point, those values are interpolated in time just as in space. Users have the option to use either linear interpolation between the two adjacent time-slices, or else simple nearest-neighbor sampling. The latter can be useful for simulation dumps that are not too finely sampled ($\Delta t \gtrsim 1\ G M / c^3$), as linear interpolation on timescales longer than the magnetic field coherence time tends to systematically underestimate the magnitude of the field \citep{Dexter2010}.

\begin{figure}
  \centering
  \includegraphics{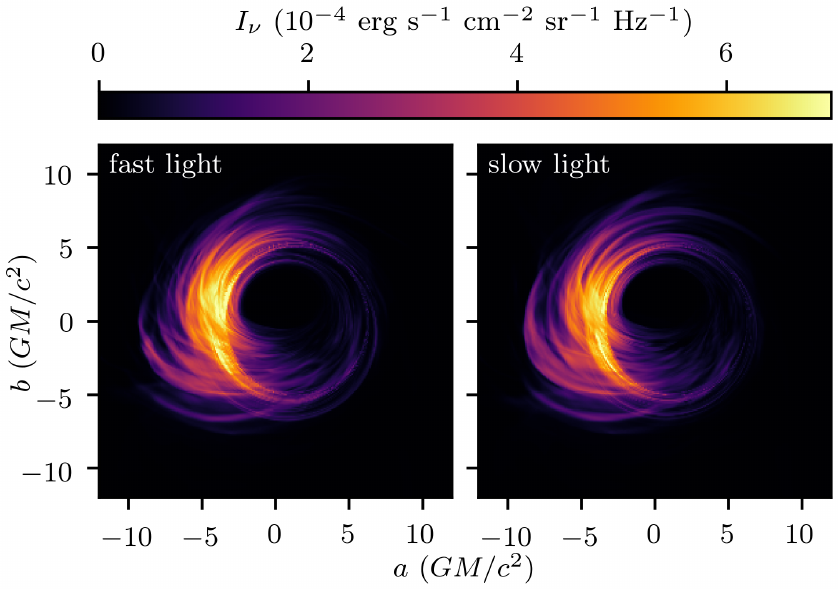}
  \caption{Results of a fast-light calculation performed with a snapshot from the fiducial simulation (left), together with a comparable slow-light calculation using $160$ snapshots spaced by $\Delta t = 1\ G M / c^3$. Though the images largely agree, the time-variability properties of the light curves generated from sequences of such images could well be different. \label{fig:slow}}
\end{figure}

Figure~\ref{fig:slow} shows the intensity from a fast-light calculation, together with a comparable slow-light image. The fast-light case is the fiducial simulation at time $10{,}500\ G M / c^3$, though the camera center (and thus outer cutoff for data being used) is moved in to $r = 25\ G M / c^2$. Even with such a small spatial extent, the longest-duration geodesics need $159\ G M / c^3$ to propagate from their termination points to the camera. The slow-light calculation uses $160$ data dumps, equally spaced from $10{,}371$ to $10{,}530\ G M / c^3$. The offset in this latter time (the time at the camera) from the fast-light time roughly accounts for the light-travel time for the bulk of the emission to the camera. The two images are quite similar, though the approaching plasma on the left side of the image is able to emit light along a number of rays at different times as it moves relativistically.

\subsection{Adaptive Ray Tracing}
\label{sec:features:adaptive}

It is sometimes desirable to use a large number of rays (i.e., pixels) in one region of an image, resolving small-scale features, while not wasting resources in other parts of the image. More codes are adopting adaptive ray-tracing techniques in order to facilitate more efficient computation; \citet{Wong2021} refines regions with particularly long geodesic path lengths, \citet{Gelles2021} use an estimate of the error derived from nearby pixels' intensities, and \citet{Davelaar2021} use a scheme very similar to what we describe here in order to trace through a binary black hole system.

\begin{figure}
  \centering
  \includegraphics{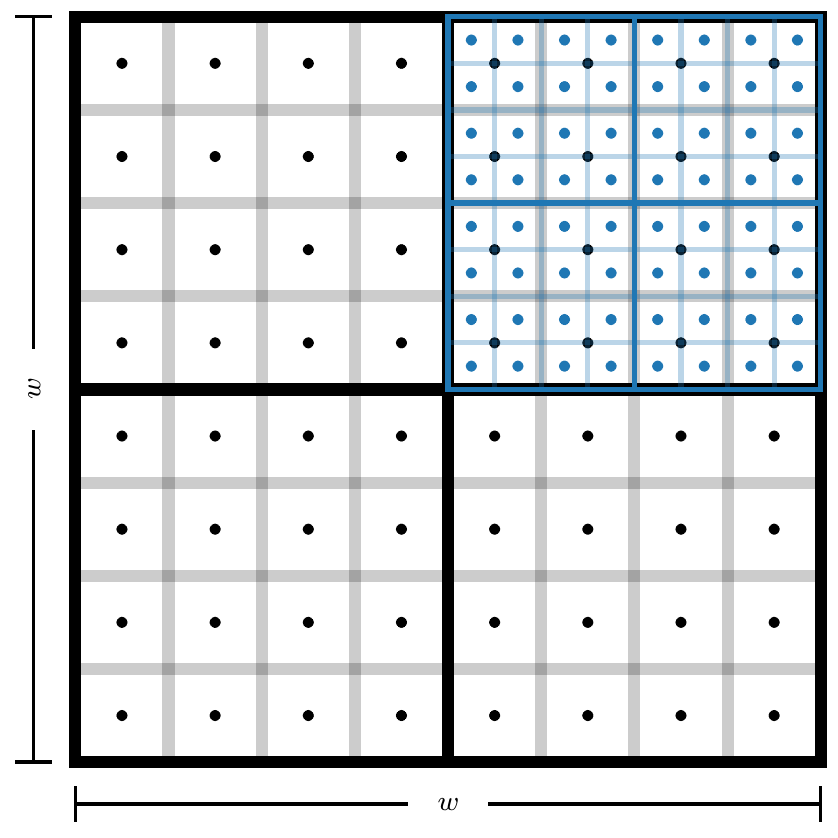}
  \caption{Schematic illustrating the arrangement of pixels into blocks suitable for refinement. The root grid samples $8^2$ rays (black points) in as many pixels (gray squares), which are arranged into four blocks of $4^2$ pixels (black squares). The upper-right block is refined into four new blocks (solid blue squares) of $4^2$ pixels each (pale blue squares), sampled at their respective centers (blue points). \label{fig:adaptive_grid}}
\end{figure}

The adaptivity in \code{Blacklight} is reminiscent of mesh refinement in \code{Athena++}. Rather than considering a pixel to be a point, the code considers a pixel to be a square region of the image plane, sampled by a ray placed at its center. This is illustrated in Figure~\ref{fig:adaptive_grid}. It is a simple matter to recursively subdivide pixels into four, with each iteration augmenting a given ray with four new rays. The root grid, which we will number adaptive level $0$, is domain-decomposed into an integer number of square blocks of pixels, whose size is specified by the user. Each block is analyzed as a whole, though with no information from other blocks, as the code runs. If a refinement condition is triggered, every pixel in the block is divided into four, and the four new blocks covering the region, considered to be at the next adaptive level, are ray-traced and evaluated for further refinement. This process continues up to a user-specified maximum refinement level, and there are no limits to how rapidly refinement can change across the image plane. The code supports single-pixel blocks, allowing fine-grained control over which regions are refined.

Blacklight supports two types of refinement criteria. One, similar to static mesh refinement in \code{Athena++} and other hydrodynamical codes, specifies regions in the image plane and corresponding minimum refinement levels. Any block whose center lies within a region and whose level is less than the given minimum will be refined. The other class of criteria is based on evaluating the set of intensity values $I_\nu^{i,j}$ ($i$ and $j$ indexing pixel column and row) at a chosen frequency across all pixels in the block. Users can specify thresholds in intensity $I_\nu^{i,j}$ itself; absolute pixel-to-pixel intensity gradient
\begin{subequations} \begin{align}
  \abs{\nabla_\mathrm{abs} I_\nu}^{i,j} & = \parenpow{\big}{\Delta_i^2 + \Delta_j^2}{1/2}, \\
  \Delta_i & = \frac{1}{2} (I_\nu^{i+1,j} - I_\nu^{i-1,j}), \\
  \Delta_j & = \frac{1}{2} (I_\nu^{i,j+1} - I_\nu^{i,j-1});
\end{align} \end{subequations}
relative pixel-to-pixel intensity gradient
\begin{subequations} \begin{align}
  \abs{\nabla_\mathrm{rel} I_\nu}^{i,j} & = \parenpow{\big}{\Delta_i^2 + \Delta_j^2}{1/2}, \\
  \Delta_i & = \frac{2 (I_\nu^{i+1,j} - I_\nu^{i-1,j})}{I_\nu^{i-1,j} + 2 I_\nu^{i,j} + I_\nu^{i+1,j}}, \\
  \Delta_j & = \frac{2 (I_\nu^{i,j+1} - I_\nu^{i,j-1})}{I_\nu^{i,j-1} + 2 I_\nu^{i,j} + I_\nu^{i,j+1}};
\end{align} \end{subequations}
absolute pixel-to-pixel Laplacian
\begin{subequations} \begin{align}
  \abs{\nabla_\mathrm{abs}^2 I_\nu}^{i,j} & = \abs{\Delta_i + \Delta_j}, \\
  \Delta_i & = I_\nu^{i-1,j} - 2 I_\nu^{i,j} + I_\nu^{i+1,j}, \\
  \Delta_j & = I_\nu^{i,j-1} - 2 I_\nu^{i,j} + I_\nu^{i,j+1};
\end{align} \end{subequations}
and/or relative pixel-to-pixel Laplacian
\begin{subequations} \begin{align}
  \abs{\nabla_\mathrm{rel}^2 I_\nu}^{i,j} & = \abs{\Delta_i + \Delta_j}, \\
  \Delta_i & = \frac{4 (I_\nu^{i-1,j} - 2 I_\nu^{i,j} + I_\nu^{i+1,j})}{I_\nu^{i-1,j} - 2 I_\nu^{i,j} + I_\nu^{i+1,j}}, \\
  \Delta_j & = \frac{4 (I_\nu^{i,j-1} - 2 I_\nu^{i,j} + I_\nu^{i,j+1})}{I_\nu^{i,j-1} - 2 I_\nu^{i,j} + I_\nu^{i,j+1}}.
\end{align} \end{subequations}
One-sided differences are substituted as appropriate near the edges of blocks. If a user-specified critical fraction of pixels in a block surpasses a given threshold, the block is refined.

\begin{figure*}
  \centering
  \includegraphics{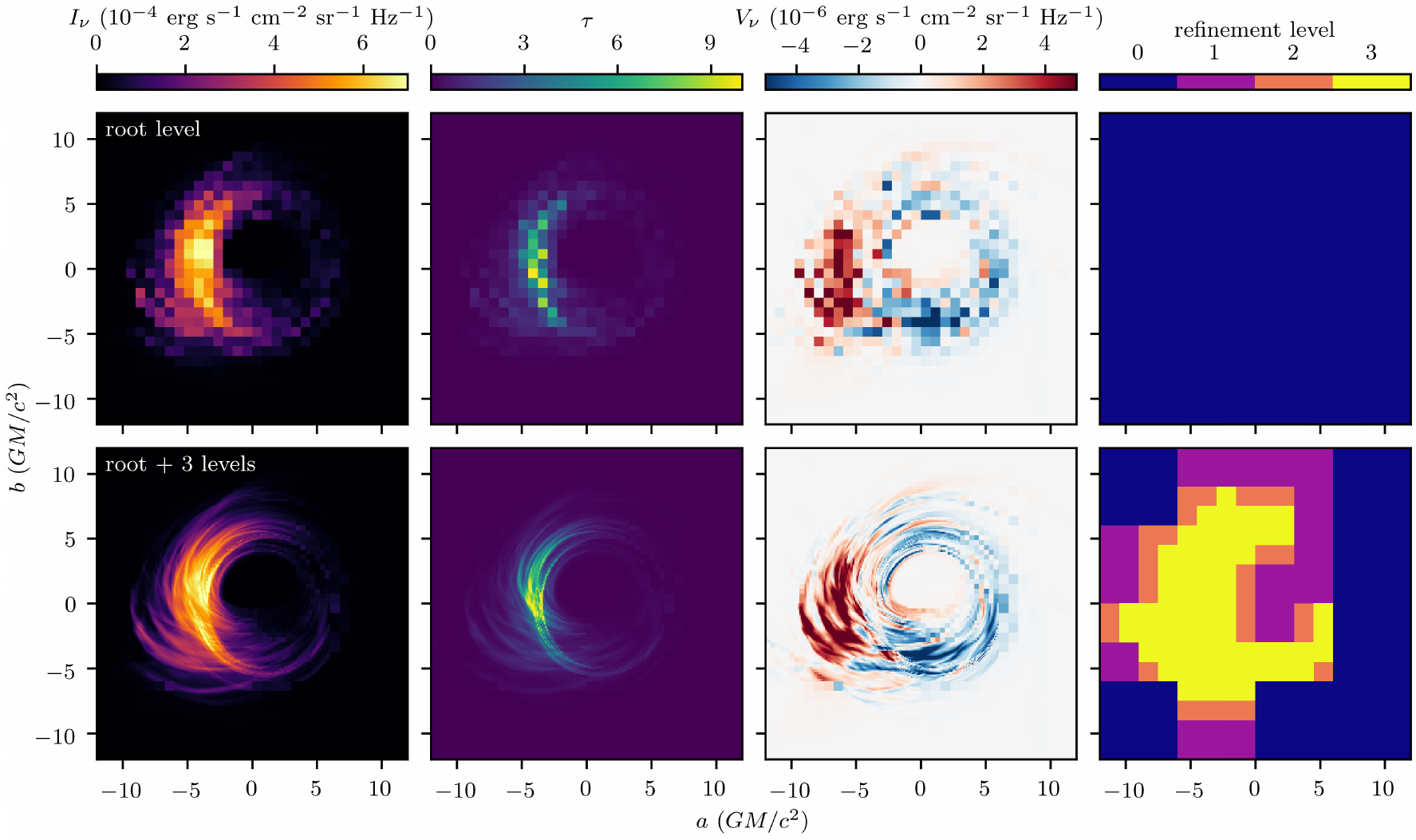}
  \caption{Unrefined (top) and adaptively refined (bottom) images of the fiducial simulation. The root level consists of $32^2$ pixels in $8^2$ blocks, and refinement is controlled via the absolute pixel-to-pixel intensity gradient. Refinement successfully captures interesting structure in intensity, optical depth, and circular polarization, without increasing resolution in the outer parts of the image. \label{fig:adaptive}}
\end{figure*}

As an illustration, we consider the fiducial simulation with a root resolution of $32^2$ pixels, broken into blocks of $8^2$ pixels. The top panels of Figure~\ref{fig:adaptive} show the intensity, optical depth, and circular polarization at root level. Though rapidly computed, these images fail to show any features beyond the presence of a crescent with a spur. We then add up to three levels of refinement, refining any blocks for which $\abs{\nabla_\mathrm{abs} I_\nu} > 5 \times 10^{-5}\ \erg\ \s^{-1}\ \cm^{-2}\ \sr^{-1}\ \hz^{-1}$ for five or more pixels. As shown in the lower panels, this criterion focuses pixels in the bright parts of the image with plentiful features, keeping resolution low in the extremities of the image and in the darker parts of the crescent.

\subsection{Nonthermal Electrons}
\label{sec:features:nonthermal}

As discussed in \S\ref{sec:integration:coefficients}, \code{Blacklight} supports thermal, power-law, and kappa distributions for the electrons. A majority of literature results to date assume purely thermal electrons, though this is largely a de facto choice that should become less common as small-scale plasma research continues to inform the GRMHD community about the properties we expect the electrons to have. As more observations are conducted, especially outside the millimeter, alternative distributions may become more necessary to match what is seen in nature.

\begin{figure}
  \centering
  \includegraphics{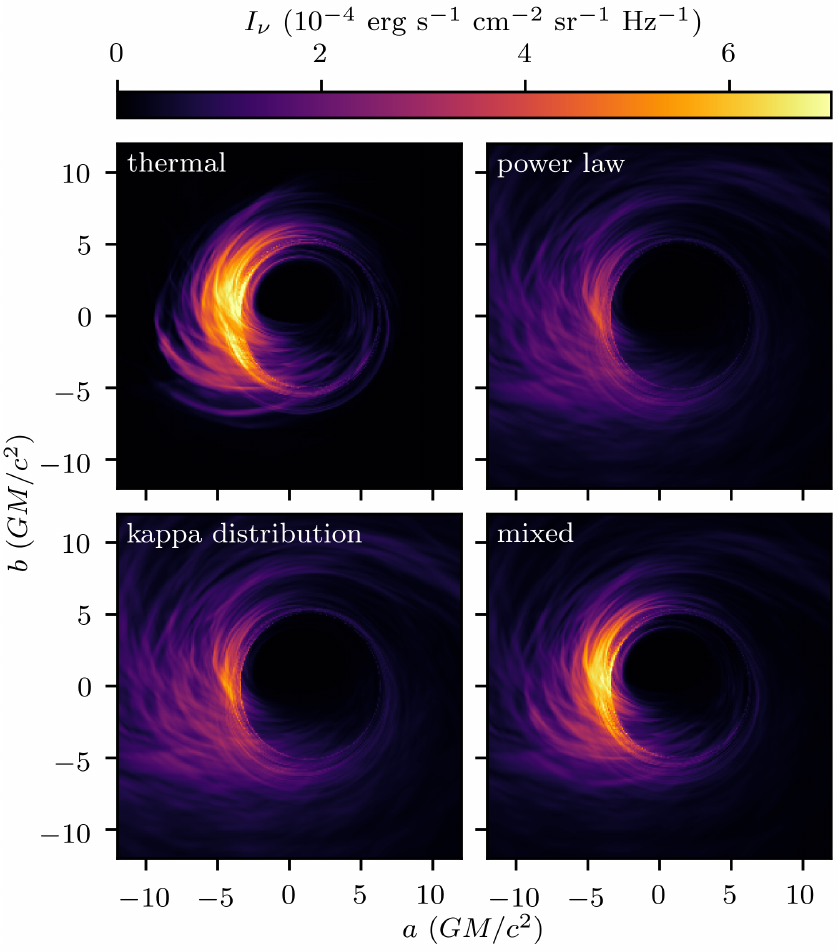}
  \caption{The fiducial image resulting from thermal electrons, and three variations made with alternate electron distributions. The power-law and kappa-distribution parameters are chosen to match the total flux of the fiducial image. The ``mixed'' case assumes one-third of the electrons are in each of the other cases. \label{fig:nonthermal}}
\end{figure}

Figure~\ref{fig:nonthermal} shows the fiducial image, made with a thermal distribution whose temperature varies across the simulation, together with nonthermal images. The power law is taken to have $p = 3$, $\gammamin = 4$, and $\gammamax = 1000$; we parameterize the kappa distribution with $\kappa = 4$ and $w = 1 / 2$. These parameters result in total fluxes comparable to the thermal case, though the images are more extended and diffuse. The lower-right panel shows the results from declaring that the electrons are equally partitioned among the three cases, achieved by simply setting the relative fractions in an input file. While not physically motivated, this demonstrates how the code can be used to quickly investigate the effects of more complicated scenarios, such as adding some power-law electrons to a mostly thermal population.

\subsection{Auxiliary Images}
\label{sec:features:auxiliary}

One of the primary goals of \code{Blacklight} is facilitating science by connecting the physical processes in a simulation with the resulting observables. One way this is accomplished is by providing a number of auxiliary ``images'' of quantities other than Stokes parameters. This allows for quick access to properties of geodesics and representative simulation values along those geodesics as functions of image-plane location.

\begin{figure}
  \centering
  \includegraphics{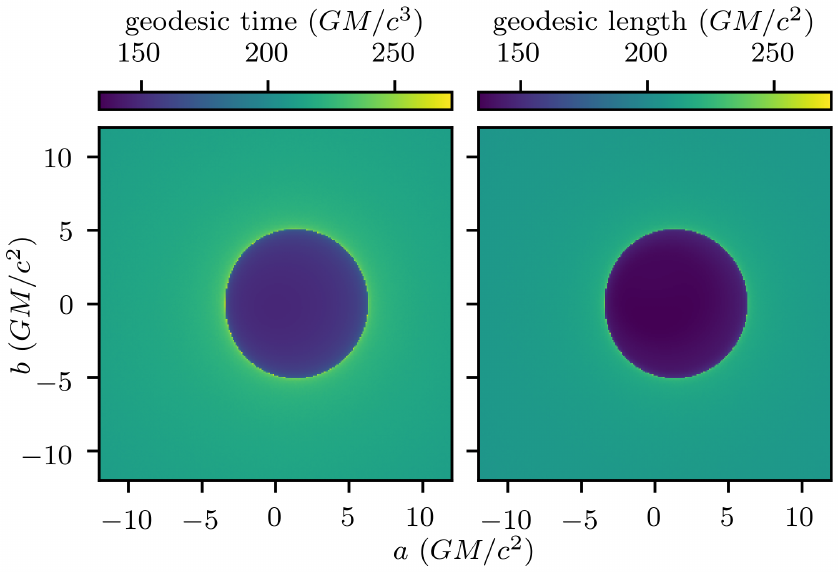}
  \caption{Elapsed time and proper length of geodesics in the fiducial simulation. The color scales are the same in geometric units, meaning the slight color differences reflect the fact that $\mathrm{d}s / \mathrm{d}t$ is not necessarily unity (the speed of light) for null geodesics in a curved spacetime. \label{fig:auxiliary_length}}
\end{figure}

One pair of simple parameters is the time and length of each geodesic. The time here is the difference in Kerr--Schild $t$ between the origin of the geodesic (its termination point when tracing backward) and the camera. For length, \code{Blacklight} reports the Kerr--Schild proper length $s$ obtained by integrating \eqref{eq:length} along the geodesic. Figure~\ref{fig:auxiliary_length} shows these two quantities for the fiducial simulation. The time is particularly useful for gauging how many snapshots are required to perform a slow-light calculation with a given camera.

\begin{figure*}
  \centering
  \includegraphics{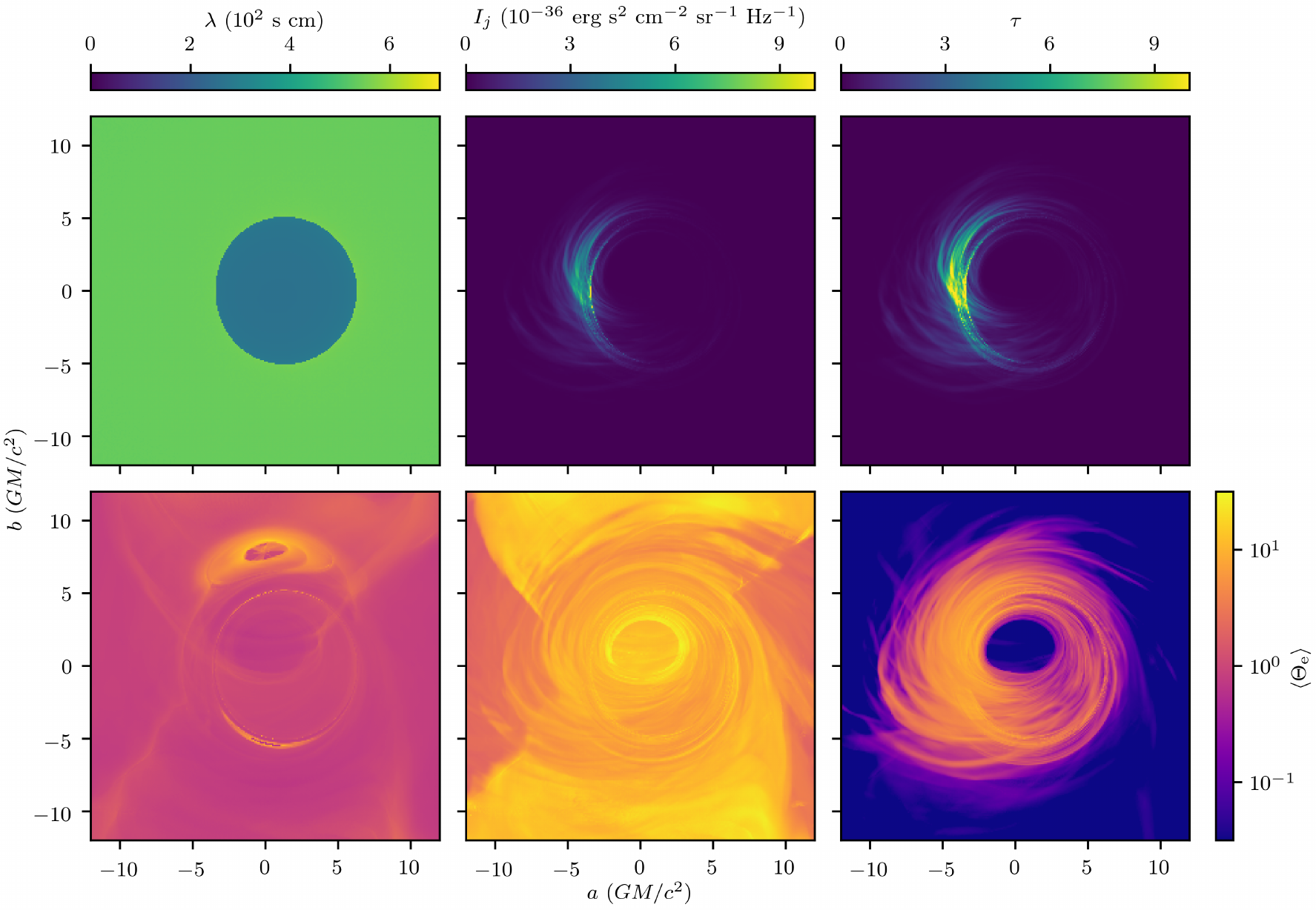}
  \caption{Top:\ auxiliary images of affine parameter, unpolarized emission, and unpolarized optical depth in the fiducial simulation. Bottom:\ averages of electron temperature weighted by the scheme corresponding to the above quantity, as described in the text. Each weighting emphasizes different parts of the simulation. \label{fig:auxiliary_average}}
\end{figure*}

We consider three additional quantities indicative of some total property of each geodesic. The code can provide images of the elapsed affine parameter $\lambda$, scaled arbitrarily as defined in \S\ref{sec:integration:camera}; it can report the integrated unpolarized emissivity
\begin{equation}
  I_j = \int j_I \, \dd\lambda;
\end{equation}
and it can calculate the total unpolarized absorption optical depth,
\begin{equation}
  \tau = \int \alpha_I \, \dd\lambda.
\end{equation}
These three quantities are displayed in the top panels of Figure~\ref{fig:auxiliary_average}.

More useful are averages of simulation quantities motivated by these three variables. As discussed in reference to data cuts (\S\ref{sec:features:cuts}), \code{Blacklight} tracks the values of $\rho$, $\nne$, $p$, $\thetae$, $B$, $\sigma$, and $\beta^{-1}$. For each such quantity $q$, it can report an affine-parameter-averaged value
\begin{equation}
  \ave{q}_\lambda = \frac{1}{\lambda} \int q \, \dd\lambda,
\end{equation}
which weights the quantity relatively uniformly over the geodesic. Alternatively, the emission-weighted average
\begin{equation}
  \ave{q}_{I_j} = \frac{1}{I_j} \int q j_I \, \dd\lambda
\end{equation}
emphasizes those parts of the ray in which the most light is being emitted. Quantities similar to this have already been put to use in the literature. However, emission weighting can be deceptive when a bright emitting region is masked by a cooler, optically thick part of the simulation, in which case the average is not representative of the conditions where the observed photons are emitted. For this reason, we introduce an optical-depth-integrated value $\ave{q}_\tau$, calculated as the solution to
\begin{equation} \label{eq:tau_int}
  \deriv{\ave{q}_\tau}{\tau} = q - \ave{q}_\tau
\end{equation}
in the same way that $I$ is the solution to the unpolarized radiative transfer equation \eqref{eq:unpolarized}. As the ray is integrated toward the camera, $\ave{q}_\tau$ is pulled toward the local value of $q$ at a rate proportional to the absorptivity $\alpha_I$. The values of $\ave{\thetae}_\lambda$, $\ave{\thetae}_{I_j}$, and $\ave{\thetae}_\tau$ are shown in the lower panels of Figure~\ref{fig:auxiliary_average}.

\subsection{False-Color Renderings}
\label{sec:features:renderings}

The auxiliary images of \S\ref{sec:features:auxiliary} all correspond to a single scalar value generated for each pixel in the image, not unlike true intensity images. \code{Blacklight} supports another type of image, which we call a false-color rendering, in which each ray has a human-perceivable color that is modified as it propagates through the simulation.

Consider a quantity $q_1$, drawn from the same set available to data cuts (\S\ref{sec:features:cuts}) and auxiliary images (\S\ref{sec:features:auxiliary}):\ $\{\rho,\, \nne,\, p,\, \thetae,\, B,\, \sigma,\, \beta^{-1}\}$. Associate to this quantity a color $c_1$; an optical depth per unit length $\tau_{s,1}$; and threshold values $q_{1,-}$ and $q_{1,+}$, either of which may be infinite. A ray, initialized to black, can have its color $c$ be modified in much the same way as the optical-depth integration \eqref{eq:tau_int},
\begin{equation}
  \deriv{c}{\tau_1} = c_1 - c,
\end{equation}
where here the appropriate optical depth is related to proper distance $s$ via
\begin{equation}
  \tau_1 = \begin{cases}
    s \tau_{s,1}, & q_{1,-} < q < q_{1,+}; \\
    0, & \text{otherwise.}
  \end{cases}
\end{equation}
In this way, the ray's color approaches $c_1$ whenever it passes through regions of spacetime where $q$ is between its cutoff values.

Alternatively, let quantity $q_2$ have color $c_2$, opacity $\alpha_2$, and threshold $q_{2,0}$. Then whenever the ray crosses a surface $q_2 = q_{2,0}$ (possibly only in a user-specified direction), we can blend its color with $c_2$ according to
\begin{equation}
  c \leftarrow (1 - \alpha_2) c + \alpha_2 c_2.
\end{equation}

Renderings defined in this way can reveal three-dimensional structure in simulation data. Note that these are more akin to volume renderings in scientific visualization software, rather than externally illuminated isosurface renderings.

We pause to note that it is all too common to find color blending schemes that work in RGB color space. However, the mixing of colored light being emulated by the above equations will fail to match human color perception if done componentwise in this space, obfuscating the scientific accuracy of the final image. Internally, \code{Blacklight} uses the correct CIE~XYZ color space designed for this purpose \citep{Smith1932}, which has the added benefit of having a gamut that includes all colors visible to the human eye. Only the final combined image, generated from an arbitrary number of quantities $q_i$ and their associated parameters, need be projected into RGB space.\footnote{Even this last projection is of course only necessary for producing standard-format digital images. Colors used for physical printing or further manipulation should utilize the XYZ data directly.}

\begin{figure*}
  \centering
  \includegraphics{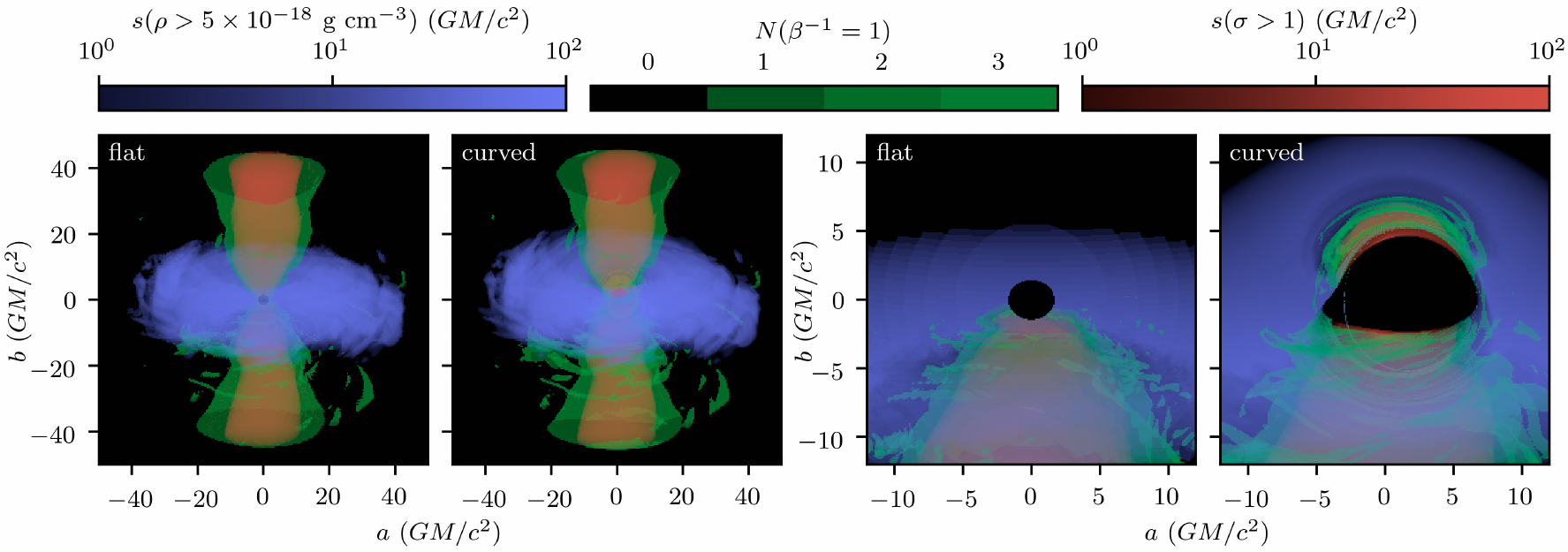}
  \caption{False-color renderings of the fiducial simulation, highlighting high densities $\rho$ (blue), $\beta^{-1} = 1$ surfaces (green), and large magnetizations $\sigma$ (red). For $\rho$ and $\sigma$, the color bars indicate the proper distance needed for a ray to achieve a given color; the color bar for $\beta$ counts how many surfaces the ray crosses. The zoomed-out images use flat spacetime (first panel) and the appropriate Kerr spacetime (second panel) for ray tracing. For the zoomed-in images (third panel, flat; fourth panel, curved), only matter with $z < -\rhor$ is considered. \label{fig:render_edge}}
\end{figure*}

Figure~\ref{fig:render_edge} shows an example of a three-component rendering from the fiducial simulation. The camera is placed close to the midplane ($\theta = 70^\circ$), and the field of view is enlarged for the first set of images. Blue traces regions with $\rho > 5 \times 10^{-18}\ \g\ \cm^{-3}$, with $\tau_s = (30\ G M / c^2)^{-1}$; red traces regions with $\sigma > 1$ (the standard $\sigma$ cut has been removed), with $\tau_s = (25\ G M / c^2)^{-1}$; and green tracks $\beta^{-1} = 1$ surfaces, with $\alpha = 0.25$. Material outside $r = 45\ G M / c^2$ is ignored here.

The first panel, made assuming flat spacetime, is akin to what general-purpose scientific visualization software would produce. This shows the spatial structure of the simulation data in an intuitive way. The second panel more directly connects the simulation variables to lensed images, as it is made with rays that follow the actual null geodesics of the spacetime. The last two panels illustrate the same dichotomy, but zoomed in to the black hole, with material outside $r = 20\ G M / c^2$ excluded. Additionally, for these latter renderings, we exclude all material with coordinate $z > -\rhor$. With this cut, we can clearly tell that much of the highly magnetized plasma seen above the black hole actually resides below it.

\begin{figure*}
  \centering
  \includegraphics{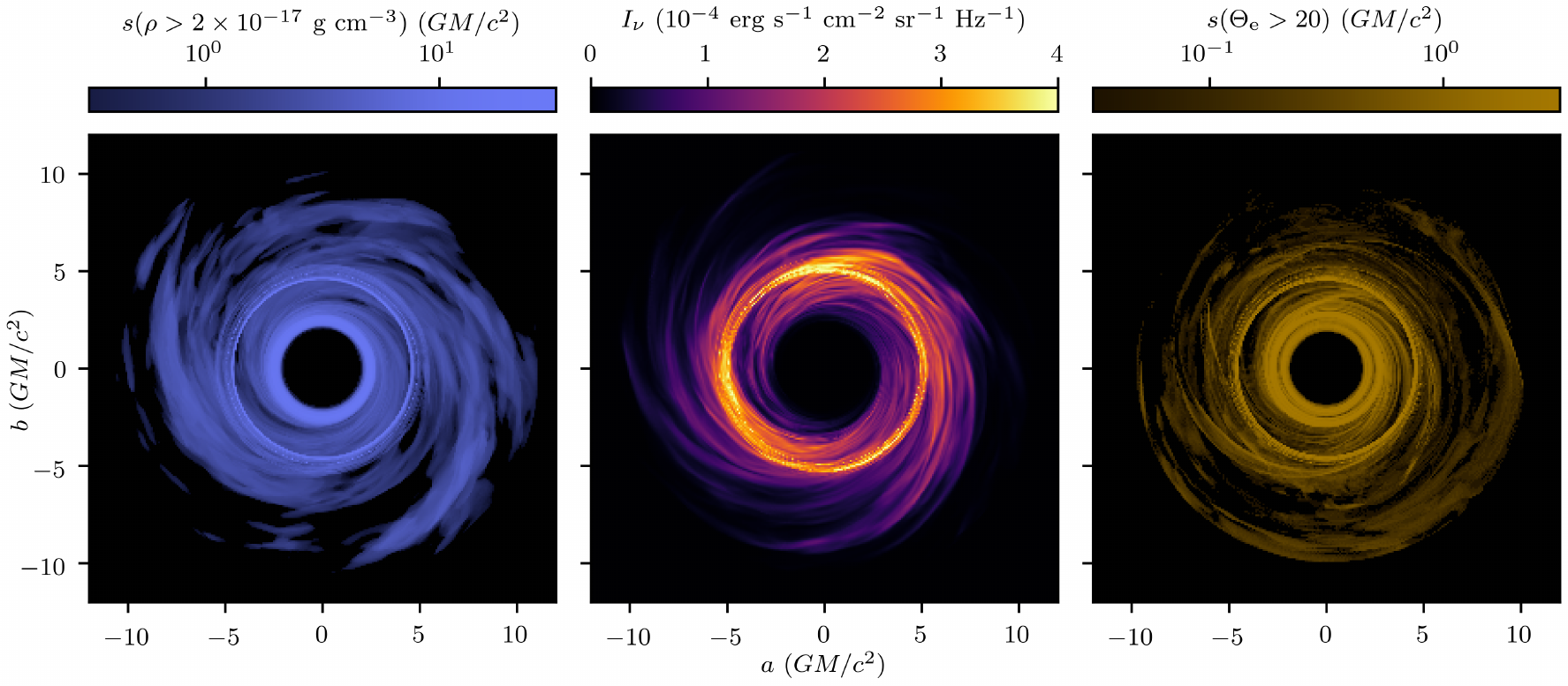}
  \caption{False-color, face-on images generated from the innermost regions of the fiducial simulation. Intensity (middle) can be compared with high densities (left) or high electron temperatures (right), with the former better matching the spiral structure easily seen by eye. For $\rho$ and $\thetae$, the color bars indicate the proper distance needed for a ray to achieve a given color. \label{fig:render_face}}
\end{figure*}

Figure~\ref{fig:render_face} provides another illustration of what can be learned from false-color renderings. The central panel shows the true $230\ \ghz$ image of the fiducial simulation as seen from the pole ($\theta = 0$), with all material outside $r = 10\ G M / c^2$ omitted. On the left, we illuminate dense ($\rho > 2 \times 10^{-17}\ \g\ \cm^{-3}$) regions, with $\tau_s = (5\ G M / c^2)^{-1}$. On the right, we illuminate regions with relativistically hot electrons ($\thetae > 20$), with $\tau_s = (1\ G M / c^2)^{-1}$. Comparisons of this sort can show at a glance what simulation variables are most correlated with image structures. Here, the spiral-armed nature of the image outside the photon ring is matched in density but not electron temperature.

\subsection{True-Color Images}
\label{sec:features:true}

The vast majority of ray-traced black hole accretion images in the literature are monochromatic, showing $I_\nu$ at a single frequency. It is, however, a simple matter to simultaneously calculate images for a number of different frequencies given a fixed camera and fixed simulation data, and there is information contained in the relationships between images at different frequencies.

Here we explore a novel means of presenting such multi-frequency data via ``true-color'' images, as seen in an appropriately boosted reference frame. The resulting images are particularly useful for outreach and public engagement, where it is often not emphasized that the ubiquitous black--reddish-orange--yellowish-white color maps used in science are merely imparting eye-pleasing colors on what is essentially grayscale information.

Given values $I_{\nu,i}$ at multiple frequencies $\nu_i$, we can self-consistently Lorentz-boost the intensity into the human visible range. Suppose we want frequency $\nu_0$ to boost to wavelength $\lambda_0'$ in a new frame. Then the intensities transform as
\begin{equation}
  I_{\lambda,i}' = c^4 \parenpow{\big}{\nu_0 \lambda_0'}{-5} \nu_i^2 I_{\nu,i}.
\end{equation}
Given $I_\lambda$ at sample wavelengths, it is straightforward to integrate against tabulated matching functions \citep{Stockman1999,Sharpe2005,Sharpe2011} based on modern measured cone response functions \citep{Stockman2000}, yielding the XYZ color coordinates of the spectrum. A standard transformation can project the resulting pixel color into RGB values. The only remaining free parameter is the overall normalization of the XYZ values, which changes the intensity but not the hue of the color. It is essentially (the reciprocal of) the camera's exposure.

\begin{figure*}
  \centering
  \includegraphics{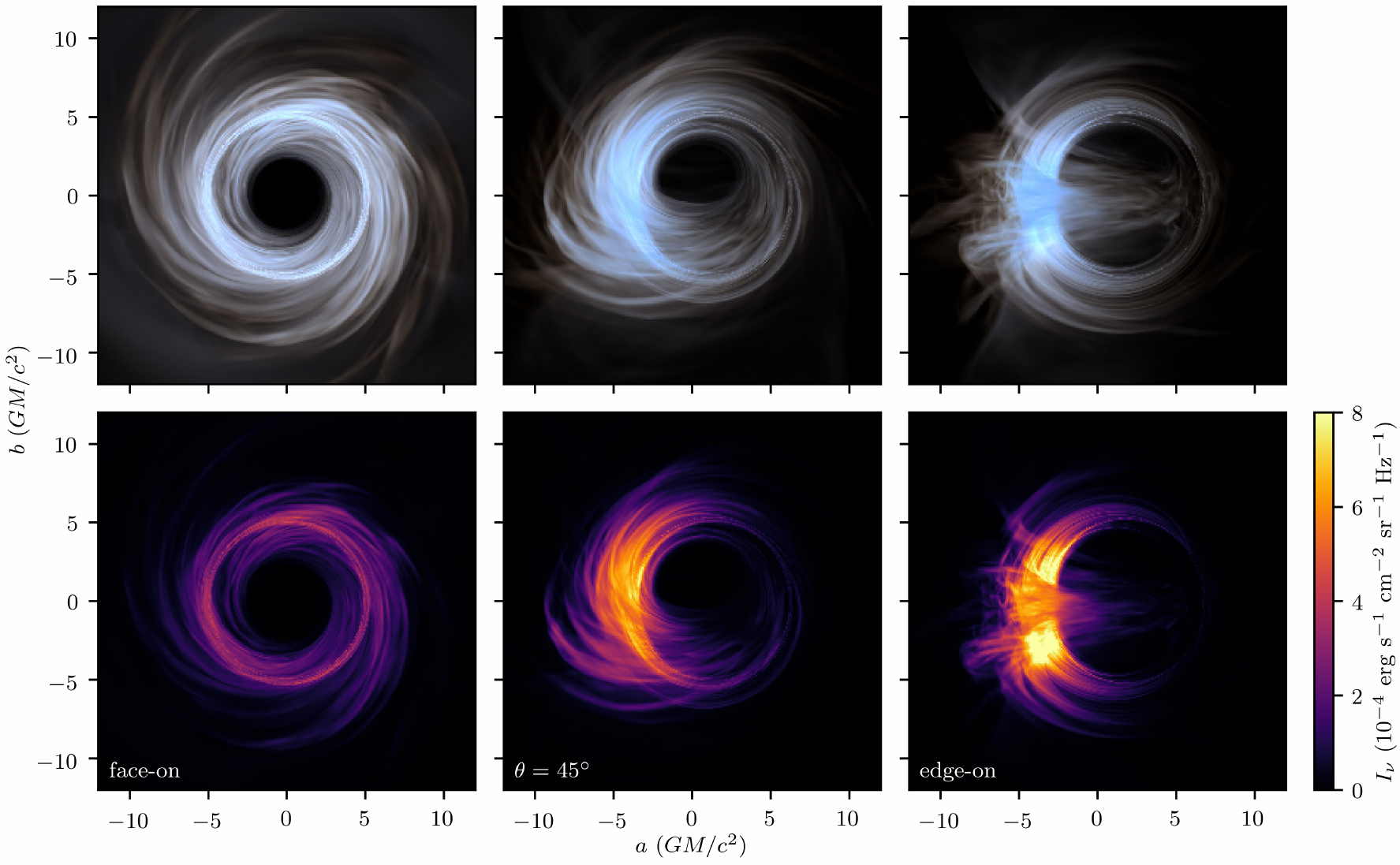}
  \caption{Top:\ true-color images of the fiducial simulation at three different viewing angles, obtained by blueshifting multi-frequency data with a Lorentz factor of approximately $1200$. Light from the left sides of the middle and right panels is emitted by plasma traveling toward the camera, and the resulting Doppler boost can be seen in the bluer color of the light. Bottom:\ the corresponding $230\ \ghz$ monochromatic images. \label{fig:true}}
\end{figure*}

We apply this procedure to the fiducial simulation, viewed face-on ($\theta = 0$), at an angle ($\theta = 45^\circ$), and edge-on ($\theta = 90^\circ$). In each case, we use $12$ frequencies, evenly spaced in wavelength from $152\ \ghz$ to $324\ \ghz$. Choosing $\nu_0 = 230\ \ghz$ and $\lambda_0' = 550\ \nm$ (near the center of human perception), the boosted wavelengths $\lambda'$ cover the visual range of $390\text{--}830\ \nm$. This corresponds to traveling toward the source at a Lorentz factor of approximately $1200$. The resulting images are shown in the upper panels of Figure~\ref{fig:true}, with the lower panels showing the standard $230\ \ghz$ monochromatic brightness maps from the same points of view.

Such boosted true-color images of synchrotron emission tend to be largely white in color. This is to be expected, as the spectrum is broad compared to the factor of ${\sim}2$ range in sensitivity of the human eye, and it tends to be monomodal. Shifting the peak frequencies to wavelengths longer or shorter than ${\sim}550\ \nm$ can make it redder or bluer, but it would be surprising to find more exotic colors as a result of this technique. Still, there are subtle color effects in Figure~\ref{fig:true}, including the fact that the approaching side of the disk is literally blueshifted in hue.

\section{Performance}
\label{sec:performance}

Though creating a single ray-traced image is much less expensive than running a GRMHD simulation, or even advancing such a simulation from one snapshot to the next, a thorough analysis based on imaging can come at considerable cost given the number of parameters such as camera position, frequency, and electron model that can be varied. We therefore provide performance numbers for \code{Blacklight},\footnote{All measurements use version 1.0 of the code (commit \code{1ab5bf2e} on 2022\nobreakdash-05\nobreakdash-22).} in order to estimate how costly any particular analysis might be.

Here we consider the polarized ray tracing of a single GRMHD snapshot, with no slow light, adaptivity, auxiliary images, or false-color renderings. The snapshot consists of a $77 \times 64 \times 128$ spherical grid. Tests are run on four different architectures:
\begin{itemize}
  \item \code{Stellar} is a campus cluster at Princeton University, consisting of $2.9\ \ghz$ Cascade Lake nodes (Intel Xeon Platinum 8268) with $96$ physical cores per node.
  \item \code{Tiger} is another campus cluster at Princeton University, consisting of $2.4\ \ghz$ Skylake nodes (Intel Xeon Gold 6148) with $40$ physical cores per node.
  \item The \code{Stampede2} cluster at the Texas Advanced Computing Center (TACC), available through the Extreme Science and Engineering Discovery Environment (XSEDE) program, has $2.1\ \ghz$ Skylake nodes (Intel Xeon Platinum 8160) with $48$ physical cores per node.
  \item \code{Stampede2} also has $1.4\ \ghz$ Knights Landing nodes (Intel Xeon Phi 7250) with $68$ physical cores per node.
\end{itemize}

\begin{figure}
  \centering
  \includegraphics{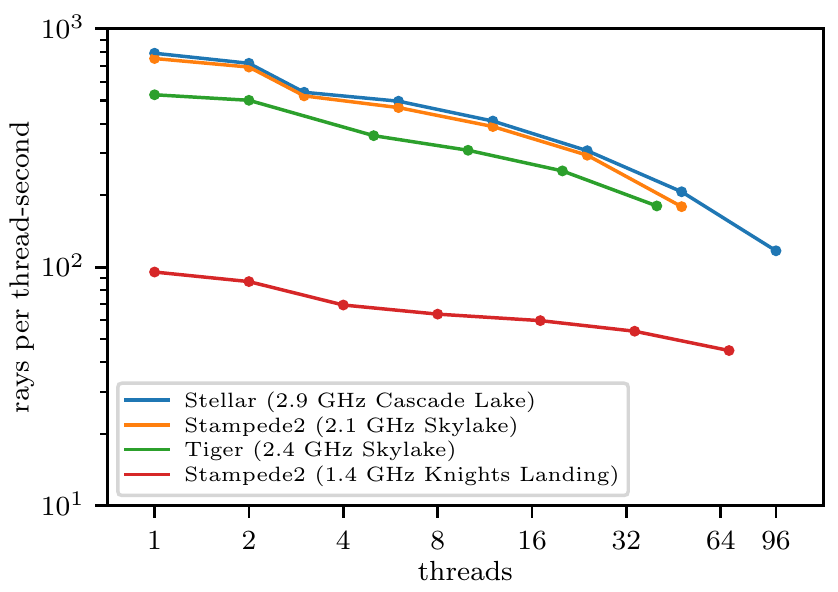}
  \caption{Parallel scaling of \code{Blacklight} with number of threads on four different machines, measured by averaging performance across multiple runs to generate each data point. The drop at high thread counts suggests that batch processing of many images should be done with multiple independent processes on a node, each with a small number of cores. \label{fig:scaling_strong}}
\end{figure}

First, we demonstrate the ``strong scaling'' of the code, fixing the image resolution to be $256^2$ pixels and measuring performance while varying the number of threads used by the code, from a single thread to one thread per physical core. The results, shown in Figure~\ref{fig:scaling_strong}, show typical values of $100\text{--}1000$ rays computed per second per thread. This includes the integration of both the geodesic equation and the radiative transfer equation, as well as the (small) amortized costs associated with bookkeeping and reading and sampling the simulation data.

There is a considerable decline in performance with increasing thread count. This is largely unrelated to the hardware itself having less capability than the number of cores might suggest, as happens when processors throttle highly parallel computations for thermal reasons or when the memory bus throughput becomes a limiting factor; the performance numbers vary little between the cases of keeping the rest of the node idle and keeping all cores busy with comparable work. The loss in efficiency is also not the result of large sections of serial code; most of the computation in \code{Blacklight} is distributed across threads via OpenMP parallel directives, where each thread works on a contiguous block of pixels. However, these parallel loops suffer performance losses with large numbers of threads, likely due to cache coherency issues. Different threads simultaneously try to read from and write to nearby (but not identical) locations in memory as they work on the parts of various arrays associated with their block of pixels.

This scaling, however, essentially only affects users wanting to use an entire node to rapidly compute a single image at very high resolution. The more common scientific use case is that of computing many images at more modest resolution (e.g., $128^2$ or $256^2$), where multiple instances of the code can fit within the memory available on a single node. In this case ${\sim}4$ cores can be assigned to each of ${\sim}10\text{--}30$ processes, with each process launching as many threads as cores. The result will be the performance seen toward the left side of Figure~\ref{fig:scaling_strong}, even with the entire node dedicated to ray tracing.

\begin{figure}
  \centering
  \includegraphics{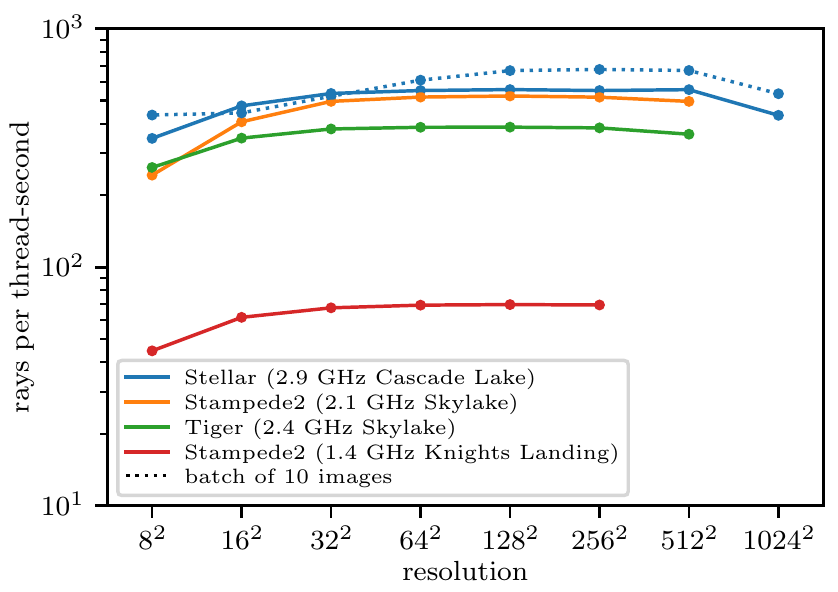}
  \caption{\code{Blacklight} performance as a function of output image resolution on four different machines, measured by averaging the results of multiple runs to generate each data point. Four threads are used in all cases. Solid lines indicate a single image is calculated with each call of the code; the dashed line corresponds to having \code{Blacklight} trace $10$ images simultaneously. \label{fig:scaling_resolution}}
\end{figure}

Figure~\ref{fig:scaling_resolution} shows how image resolution is not a factor in performance as measured in cost per pixel. Here we fix all runs to use four threads, varying the resolution of the output image. Image sizes are limited by the memory available to each node. In one case, \code{Blacklight} is used to trace $10$ images at once, rather than just one, using identical camera settings and input files with identical layouts. There are some benefits to efficiency by amortizing over many images the one-time cost of integrating the geodesic equation.

Taking $256^2$ pixels using four threads on \code{Stellar} as a representative case, each image takes $30\ \s$ to run. Of this time, $75\%$ is spent integrating the polarized radiative transfer equation, including calculating synchrotron coefficients; $13\%$ is spent resampling values from the simulation grid onto geodesics; $12\%$ is spent integrating the geodesic equation; and less than $1\%$ is spent on all other tasks. Relative to the same $30\ \s$ total, integrating the unpolarized transfer equation takes $15\%$ of the time, making the total unpolarized runtime $12\ \s$, $39\%$ that of the polarized case. If we instead produce only the $26$ auxiliary images described in \S\ref{sec:features:auxiliary}, integrating the image takes $29\%$ of $30\ \s$, resulting in a runtime of $16\ \s$, $54\%$ that of the polarized case. Finally, if we only produce a single false-color rendering with three components, as in Figure~\ref{fig:render_edge}, the integration takes $14\%$ of $30\ \s$, leading to a total runtime of $11\ \s$, $38\%$ that of the polarized case.

Further optimization of \code{Blacklight} is underway, with the goal of achieving near-perfect scaling, as should be possible given the embarrassingly parallel nature of ray tracing many pixels. Additionally, profiling tools indicate parts of the code are not yet fully vectorized (compiled to single instruction, multiple data (SIMD) instructions). We note that the absolute performance is still competitive with other codes. Using a $288 \times 128 \times 128$ \code{iharm3D} simulation data dump, we trace a $256^2$ polarized image with both \code{Blacklight} and \code{ipole} on \code{Stellar}, using \code{icpc}~2021.1.2 to compile the former and \code{gcc}~8.5.0 for the latter.\footnote{\code{ipole} additionally calculates an unpolarized image, though this should be only a small fraction of the total cost.} Though \code{ipole} has better scaling with number of threads, \code{Blacklight} calculates and processes sample points along rays faster when using fewer than $48$ threads. With a single thread, \code{Blacklight} processes $3.4 \times 10^5$ samples per second, compared to $1.0 \times 10^5$ for \code{ipole}.

\section{Code Philosophy}
\label{sec:philosophy}

The aiding of scientific investigations is the primary motivation for \code{Blacklight}, and this reflected in several aspects of the code.

Given that most academic researchers have access to disparate hardware and software environments with little direct technical support, it is important that codes for researchers work with minimal manual configuration. \code{Blacklight} has no dependencies on external libraries, placing the burden on the developer rather than the user to ensure it can interface with HDF5 and NumPy files, for example. It requires no third-party configuration tools. The aim is to have code that compiles on most Unix-like machines with no effort. The code is written in standard \code{C++17}, which is supported by all major compiler vendors.

Users of scientific codes should be able to discern the exact algorithms being employed in the source code, perhaps modifying the methods to suit their needs. Thorough documentation certainly helps, and the wiki for \code{Blacklight} is approximately one-third the size of the source code proper. Ideally, users should be able to inspect the source code and quickly find and understand pertinent sections. To this end, a moderate amount of encapsulation is employed (the code has just five main classes, for reading an input file, reading simulation data, integrating geodesics, integrating radiation, and outputting results), enough to allow users to understand parts of the code without understanding the entire code base, but not so much as would result in chasing the logic of a small task through many files. \code{Blacklight} uses only select features of \code{C++} beyond what is contained in \code{C}; it is largely imperative in style, as is appropriate for implementing straightforward algorithms that employ long sequences of mathematical equations.

Finally, \code{Blacklight} is publicly available. In fact, it is placed into the public domain. There is only the one version of the code as described here; users need not worry that an official, private, better version exists. The main repository is meant to be selective in what it incorporates, so that users can have confidence that updating to the latest version will not reduce functionality or accuracy in their own workflow.

It is hoped that these precepts will lead to a code that avoids the pitfalls common to academic tools for scientific computation. \code{Blacklight} can then assist and enable black hole accretion research, especially in the arena of understanding physical processes by connecting astrophysical simulations to astronomical observations.

\acknowledgments

We are indebted to those who led the way in providing the GRMHD community with ray-tracing codes, especially C.~Gammie, J.~Dexter, M.~Mo\'scibrodzka, and B.~Prather, who answered many of this author's questions about such codes; to O.~Blaes for extensive discussion regarding the nuances of synchrotron radiation; and to B.~Oyang, for providing the initial motivation to write this software.

This work used the Princeton Research Computing clusters \code{Stellar} and \code{Tiger} managed and supported by the Princeton Institute for Computational Science and Engineering (PICSciE) and the Office of Information Technology's High Performance Computing Center and Visualization Laboratory at Princeton University, as well as the Extreme Science and Engineering Discovery Environment (XSEDE) cluster \code{Stampede2} at the Texas Advanced Computing Center (TACC) through allocation AST200005.

\bibliographystyle{aasjournal}
\bibliography{references}

\begin{thebibliography}{}
\expandafter\ifx\csname natexlab\endcsname\relax\def\natexlab#1{#1}\fi
\providecommand{\url}[1]{\href{#1}{#1}}
\providecommand{\dodoi}[1]{doi:~\href{http://doi.org/#1}{\nolinkurl{#1}}}
\providecommand{\doeprint}[1]{\href{http://ascl.net/#1}{\nolinkurl{http://ascl.net/#1}}}
\providecommand{\doarXiv}[1]{\href{https://arxiv.org/abs/#1}{\nolinkurl{https://arxiv.org/abs/#1}}}

\bibitem[{Broderick \& Blandford(2003)}]{Broderick2003}
Broderick, A., \& Blandford, R. 2003, MNRAS, 342, 1280,
  \dodoi{10.1046/j.1365-8711.2003.06618.x}

\bibitem[{Broderick \& Blandford(2004)}]{Broderick2004}
---. 2004, MNRAS, 349, 994, \dodoi{10.1111/j.1365-2966.2004.07582.x}

\bibitem[{Bronzwaer {et~al.}(2018)Bronzwaer, Davelaar, Younsi,
  Mo{\'s}cibrodzka, Falcke, Kramer, \& Rezzolla}]{Bronzwaer2018}
Bronzwaer, T., Davelaar, J., Younsi, Z., {et~al.} 2018, AA, 613, A2,
  \dodoi{10.1051/0004-6361/201732149}

\bibitem[{Bronzwaer {et~al.}(2020)Bronzwaer, Younsi, Davelaar, \&
  Falcke}]{Bronzwaer2020}
Bronzwaer, T., Younsi, Z., Davelaar, J., \& Falcke, H. 2020, AA, 641, A126,
  \dodoi{10.1051/0004-6361/202038573}

\bibitem[{Chan {et~al.}(2018)Chan, Medeiros, {\"O}zel, \& Psaltis}]{Chan2018}
Chan, C., Medeiros, L., {\"O}zel, F., \& Psaltis, D. 2018, ApJ, 867, 59,
  \dodoi{10.3847/1538-4357/aadfe5}

\bibitem[{Chan {et~al.}(2013)Chan, Psaltis, \& {\"O}zel}]{Chan2013}
Chan, C., Psaltis, D., \& {\"O}zel, F. 2013, ApJ, 777, 13,
  \dodoi{10.1088/0004-637X/777/1/13}

\bibitem[{Davelaar {et~al.}(2018)Davelaar, Bronzwaer, Kok, Younsi,
  Mo{\'s}cibrodzka, \& Falcke}]{Davelaar2018}
Davelaar, J., Bronzwaer, T., Kok, D., {et~al.} 2018, ComAC, 5, 1,
  \dodoi{10.1186/s40668-018-0023-7}

\bibitem[{Davelaar \& Haiman(2021)}]{Davelaar2021}
Davelaar, J., \& Haiman, Z. 2021.
\newblock \doarXiv{2112.05828}

\bibitem[{Dexter(2016)}]{Dexter2016}
Dexter, J. 2016, MNRAS, 462, 115, \dodoi{10.1093/mnras/stw1526}

\bibitem[{Dexter \& Agol(2009)}]{Dexter2009}
Dexter, J., \& Agol, E. 2009, ApJ, 696, 1616,
  \dodoi{10.1088/0004-637X/696/2/1616}

\bibitem[{Dexter {et~al.}(2010)Dexter, Agol, Fragile, \& McKinney}]{Dexter2010}
Dexter, J., Agol, E., Fragile, P.~C., \& McKinney, J.~C. 2010, ApJ, 717, 1092,
  \dodoi{10.1088/0004-637X/717/2/1092}

\bibitem[{Doeleman {et~al.}(2008)Doeleman, Weintroub, Rogers, Plambeck, Freund,
  Tilanus, Friberg, Ziurys, Moran, Corey, Young, Smythe, Titus, Marrone,
  Cappallo, Bock, Bower, Chamberlin, Davis, Krichbaum, Lamb, Maness, Niell,
  Roy, Strittmatter, Werthimer, Whitney, \& Woody}]{Doeleman2008}
Doeleman, S.~S., Weintroub, J., Rogers, A.~E., {et~al.} 2008, Natur, 455, 78,
  \dodoi{10.1038/nature07245}

\bibitem[{Dormand \& Prince(1980)}]{Dormand1980}
Dormand, J., \& Prince, P. 1980, JCoAM, 6, 19,
  \dodoi{10.1016/0771-050X(80)90013-3}

\bibitem[{Gammie {et~al.}(2003)Gammie, McKinney, \& T{\'o}th}]{Gammie2003}
Gammie, C.~F., McKinney, J.~C., \& T{\'o}th, G. 2003, ApJ, 589, 444,
  \dodoi{10.1086/374594}

\bibitem[{Gelles {et~al.}(2021)Gelles, Prather, Palumbo, Johnson, Wong, \&
  Georgiev}]{Gelles2021}
Gelles, Z., Prather, B., Palumbo, D., {et~al.} 2021, ApJ, 912, 39,
  \dodoi{10.3847/1538-4357/abee13}

\bibitem[{Gold {et~al.}(2020)Gold, Broderick, Younsi, Fromm, Gammie,
  Mo{\'s}cibrodzka, Pu, Bronzwaer, Davelaar, Dexter, Ball, Chan, Kawashima,
  Mizuno, Ripperda, \& {The EHT Collaboration}}]{Gold2020}
Gold, R., Broderick, A.~E., Younsi, Z., {et~al.} 2020, ApJ, 897, 148,
  \dodoi{10.3847/1538-4357/ab96c6}

\bibitem[{{GRAVITY Collaboration}(2018)}]{Gravity2018}
{GRAVITY Collaboration}. 2018, AA, 618, L10,
  \dodoi{10.1051/0004-6361/201834294}

\bibitem[{Iyer \& Hansen(2009)}]{Iyer2009}
Iyer, S., \& Hansen, E. 2009, PhRvD, 80, 124023,
  \dodoi{10.1103/PhysRevD.80.124023}

\bibitem[{James {et~al.}(2015)James, {von~Tunzelmann}, Franklin, \&
  Thorne}]{James2015}
James, O., {von~Tunzelmann}, E., Franklin, P., \& Thorne, K.~S. 2015, AmJPh,
  83, 486, \dodoi{10.1119/1.4916949}

\bibitem[{Jones \& O'Dell(1977)}]{Jones1977}
Jones, T., \& O'Dell, S. 1977, ApJ, 214, 522, \dodoi{10.1086/155278}

\bibitem[{Kawashima {et~al.}(2021)Kawashima, Ohsuga, \&
  Takahashi}]{Kawashima2021}
Kawashima, T., Ohsuga, K., \& Takahashi, H.~R. 2021.
\newblock \doarXiv{2108.05131}

\bibitem[{{Landi~Degl'Innocenti} \&
  {Landi~Degl'Innocenti}(1985)}]{LandiDeglInnocenti1985}
{Landi~Degl'Innocenti}, E., \& {Landi~Degl'Innocenti}, M. 1985, SoPh, 97, 239,
  \dodoi{10.1007/BF00165988}

\bibitem[{Marszewski {et~al.}(2021)Marszewski, Prather, Joshi, Pandya, \&
  Gammie}]{Marszewski2021}
Marszewski, A., Prather, B.~S., Joshi, A.~V., Pandya, A., \& Gammie, C.~F.
  2021, ApJ, 921, 17, \dodoi{10.3847/1538-4357/ac1b28}

\bibitem[{Mo{\'s}cibrodzka {et~al.}(2016)Mo{\'s}cibrodzka, Falcke, \&
  Shiokawa}]{Moscibrodzka2016}
Mo{\'s}cibrodzka, M., Falcke, H., \& Shiokawa, H. 2016, AA, 586, A38,
  \dodoi{10.1051/0004-6361/201526630}

\bibitem[{Mo{\'s}cibrodzka \& Gammie(2018)}]{Moscibrodzka2018}
Mo{\'s}cibrodzka, M., \& Gammie, C. 2018, MNRAS, 475, 43,
  \dodoi{10.1093/mnras/stx3162}

\bibitem[{Noble {et~al.}(2007)Noble, Leung, Gammie, \& Book}]{Noble2007}
Noble, S.~C., Leung, P.~K., Gammie, C.~F., \& Book, L.~G. 2007, CQGra, 24,
  S259, \dodoi{10.1088/0264-9381/24/12/S17}

\bibitem[{Prather {et~al.}(2021)Prather, Wong, Dhruv, Ryan, Dolence, Ressler,
  \& Gammie}]{Prather2021}
Prather, B.~S., Wong, G.~N., Dhruv, V., {et~al.} 2021, JOSS, 6, 3336,
  \dodoi{10.21105/joss.03336}

\bibitem[{Press {et~al.}(1997)Press, Teukolsky, Vetterling, \& Flannery}]{NR}
Press, W.~H., Teukolsky, S.~A., Vetterling, W.~T., \& Flannery, B.~P. 1997,
  {Numerical Recipes:\ The Art of Scientific Computing}, 3rd edn. (Cambridge
  University Press)

\bibitem[{Pu \& Broderick(2018)}]{Pu2018}
Pu, H.-Y., \& Broderick, A.~E. 2018, ApJ, 863, 148,
  \dodoi{10.3847/1538-4357/aad086}

\bibitem[{Pu {et~al.}(2016)Pu, Yun, Younsi, \& Yoon}]{Pu2016}
Pu, H.-Y., Yun, K., Younsi, Z., \& Yoon, S.-J. 2016, ApJ, 820, 105,
  \dodoi{10.3847/0004-637X/820/2/105}

\bibitem[{Ressler {et~al.}(2015)Ressler, Tchekhovskoy, Quataert, Chandra, \&
  Gammie}]{Ressler2015}
Ressler, S., Tchekhovskoy, A., Quataert, E., Chandra, M., \& Gammie, C. 2015,
  MNRAS, 454, 1848, \dodoi{10.1093/mnras/stv2084}

\bibitem[{S{\k{a}}dowski {et~al.}(2017)S{\k{a}}dowski, Wielgus, Narayan,
  Abarca, McKinney, \& Chael}]{Sadowski2017}
S{\k{a}}dowski, A., Wielgus, M., Narayan, R., {et~al.} 2017, MNRAS, 466, 705,
  \dodoi{10.1093/mnras/stw3116}

\bibitem[{Schnittman {et~al.}(2006)Schnittman, Krolik, \&
  Hawley}]{Schnittman2006}
Schnittman, J.~D., Krolik, J.~H., \& Hawley, J.~F. 2006, ApJ, 651, 1031,
  \dodoi{10.1086/507421}

\bibitem[{Shampine(1986)}]{Shampine1986}
Shampine, L.~F. 1986, MaCom, 46, 135, \dodoi{10.1090/S0025-5718-1986-0815836-3}

\bibitem[{Sharpe {et~al.}(2005)Sharpe, Stockman, Jagla, \&
  J{\"a}gle}]{Sharpe2005}
Sharpe, L.~T., Stockman, A., Jagla, W., \& J{\"a}gle, H. 2005, JV, 5, 948,
  \dodoi{10.1167/5.11.3}

\bibitem[{Sharpe {et~al.}(2011)Sharpe, Stockman, Jagla, \&
  J{\"a}gle}]{Sharpe2011}
---. 2011, CRA, 36, 42, \dodoi{10.1002/col.20602}

\bibitem[{Smith \& Guild(1932)}]{Smith1932}
Smith, T., \& Guild, J. 1932, TrOS, 33, 73, \dodoi{10.1088/1475-4878/33/3/301}

\bibitem[{Stockman \& Sharpe(2000)}]{Stockman2000}
Stockman, A., \& Sharpe, L.~T. 2000, VR, 40, 1711,
  \dodoi{10.1016/S0042-6989(00)00021-3}

\bibitem[{Stockman {et~al.}(1999)Stockman, Sharpe, \& Fach}]{Stockman1999}
Stockman, A., Sharpe, L.~T., \& Fach, C. 1999, VR, 39, 2901,
  \dodoi{10.1016/S0042-6989(98)00225-9}

\bibitem[{Stone {et~al.}(2020)Stone, Tomida, White, \& Felker}]{Stone2020}
Stone, J.~M., Tomida, K., White, C.~J., \& Felker, K.~G. 2020, ApJS, 249, 4,
  \dodoi{10.3847/1538-4365/ab929b}

\bibitem[{{The EHT Collaboration}(2019e)}]{EHT2019e}
{The EHT Collaboration}. 2019e, ApJL, 875, L5, \dodoi{10.3847/2041-8213/ab0f43}

\bibitem[{{The EHT Collaboration}(2019f)}]{EHT2019f}
---. 2019f, ApJL, 875, L6, \dodoi{10.3847/2041-8213/ab1141}

\bibitem[{{The EHT Collaboration}(2021)}]{EHT2021}
---. 2021, ApJL, 910, L13, \dodoi{10.3847/2041-8213/abe4de}

\bibitem[{{The EHT Collaboration}(2022e)}]{EHT2022e}
---. 2022e, ApJL, 930, L16, \dodoi{10.3847/2041-8213/ac6672}

\bibitem[{{The GRAVITY Collaboration}(2019)}]{Gravity2019}
{The GRAVITY Collaboration}. 2019, AA, 625, L10,
  \dodoi{10.1051/0004-6361/201935656}

\bibitem[{{Vel\'asquez-Cadavid} {et~al.}(2022){Vel\'asquez-Cadavid},
  {Arrieta-Villamizar}, {Lora-Clavijo}, Pimentel, \&
  {Osorio-Vargas}}]{VelasquezCadavid2022}
{Vel\'asquez-Cadavid}, J., {Arrieta-Villamizar}, J., {Lora-Clavijo}, F.,
  Pimentel, O., \& {Osorio-Vargas}, J. 2022, EPJC, 82

\bibitem[{Walker \& Penrose(1970)}]{Walker1970}
Walker, M., \& Penrose, R. 1970, CMaPh, 18, 265, \dodoi{10.1007/BF01649445}

\bibitem[{White {et~al.}(2020)White, Dexter, Blaes, \& Quataert}]{White2020}
White, C.~J., Dexter, J., Blaes, O., \& Quataert, E. 2020, ApJ, 894, 14,
  \dodoi{10.3847/1538-4357/ab8463}

\bibitem[{White {et~al.}(2016)White, Stone, \& Gammie}]{White2016}
White, C.~J., Stone, J.~M., \& Gammie, C.~F. 2016, ApJS, 225, 22,
  \dodoi{10.3847/0067-0049/225/2/22}

\bibitem[{Wong(2021)}]{Wong2021}
Wong, G.~N. 2021, ApJ, 909, 217, \dodoi{10.3847/1538-4357/abdd2d}

\bibitem[{Wong {et~al.}(2022)Wong, Prather, Dhruv, Ryan, Mo{\'s}cibrodzka, kwan
  Chanand, Joshi, Yarza, Ricarte, Shiokawa, Dolence, Noble, McKinney, \&
  Gammie}]{Wong2022}
Wong, G.~N., Prather, B.~S., Dhruv, V., {et~al.} 2022, ApJS, 259

\bibitem[{Younsi {et~al.}(2020)Younsi, Porth, Mizuno, Fromm, \&
  Olivares}]{Younsi2020}
Younsi, Z., Porth, O., Mizuno, Y., Fromm, C.~M., \& Olivares, H. 2020, in
  {Perseus in Sicily:\ From Black Hole to Cluster Outskirts}, ed. K.~Asada,
  E.~{de~Gouveia~Dal~Pino}, M.~Giroletti, H.~Nagai, \& R.~Nemmen, Proceedings
  IAU Symposium No. 342, Portland, OR, 9--12, \dodoi{10.1017/S1743921318007263}

\end{thebibliography}

\end{document}